\documentstyle[prl,aps,epsf,multicol]{revtex}
\begin{document}
\draft

\bibliographystyle{prsty}

\title{Charge Fluctuations in the Single Electron Box}
\author{Georg G\"oppert and Hermann Grabert}
\address{Fakult\"at f\"ur Physik, Albert-Ludwigs-Universit{\"a}t, \\
Hermann-Herder-Stra{\ss}e~3, D-79104 Freiburg, Germany}

\date{\today}
\maketitle
\widetext

\begin{abstract}
Quantum fluctuations of the charge in the single electron
box are investigated. 
Based on a diagrammatic expansion we calculate the 
average island charge number
and the effective charging energy in third order in the
tunneling conductance. Near the degeneracy point where 
the energy of two charge states coincides, the 
perturbative approach fails, and we 
explicitly resum the leading logarithmic divergencies to all orders.
The predictions for zero
temperature are compared with
Monte Carlo data and with recent renormalization group
results. While good agreement between the third order result
and numerical data
justifies the perturbative approach in most of the 
parameter regime relevant experimentally,
near the degeneracy point and at zero temperature
the resummation is shown to be insufficient
to describe strong tunneling effects quantitatively.
We also determine the charge noise spectrum
employing a projection operator technique. Former 
perturbative and semiclassical results are extended by the
approach.
\end{abstract}


\raggedcolumns                             %
\begin{multicols}{2}                       %
\narrowtext                                %

\section{Introduction}

\noindent
Tunneling of electrons in metallic nanostructures is strongly 
affected by Coulomb repulsion. 
Provided the screening
length in the metallic films is small compared to tunneling barrier 
thickness and sample size, the
Coulomb energy can be written in terms of a geometrical 
capacitance $C$. The relevant energy scale of the system is the
corresponding charging energy $E_C=e^2/2C$ 
\cite{Nato92} that is the energy 
needed to charge the capacitance $C$ by one electron. 
For temperatures well below this energy, $k_B T \ll E_C$, tunneling 
onto a metallic island is exponentially suppressed. 
For weak tunneling strength, $G_T \ll G_K$, 
where $G_T$ is the phenomenological 
tunneling conductance and $G_K=e^2/h$ the conductance quantum, systems
are well described by the perturbative approach in $\alpha= G_T/G_K$ 
\cite{Averin91,IngoldNato92in}. 
When the tunneling conductance becomes larger,
higher orders in the perturbative series in $\alpha$ such as
cotunneling \cite{AverinNato92in} have to be included. 
Even at zero temperature these processes 
are not forbidden energetically, therefore, higher order 
corrections are most 
pronounced at low temperatures where first order processes are
exponentially suppressed. 
Due to the large number of terms in the perturbative
series, the approach remains restricted to the first few 
orders in $\alpha$ and one is interested in the range of validity 
one could expect. At higher temperatures, $k_B T \gg E_C$, 
thermal fluctuations dominate and perturbation theory (PT) 
becomes more accurate.
Therefore, to give a lower bound of the validity of PT 
it is sufficient to
consider the zero temperature case where PT is worst.
In fact, PT at zero temperature even diverges 
at the degeneracy point of the single electron box (SEB) 
showing that the range of validity depends on both,  
tunneling strength and applied gate voltage.

While partial resummation techniques 
\cite{MatveevBOXJETP91,Schoeller2StatePRB94,ZaikinNCAPRB94,FalciScalePRL95}
lead to a nonperturbative finite result even at the 
degeneracy point, they
need an arbitrary cutoff that limits their use for direct comparison 
with experiments. Recently, 
renormalization group (RG) ideas \cite{KoenigRGPRL98} have been 
used to remove the cutoff yielding results that depend on
parameters measurable experimentally. On the other hand, 
a complete resummation of PT can
be achieved in phase representation leading to a formally exact 
path integral formulation of single electron devices
\cite{SchoenREP90,GeorgSemiclEPJB00}. 
Here, the phase is the conjugate operator to 
the island charge number and can be related to the physical
voltage drop across the junction. The functional
can serve as starting point for analytical predictions in the
semiclassical limit
\cite{BenJacPRL83,ZaikinECPRL91,ZaikinSJPRB92,ZaikinSETJETP96,WangBOXPRB96,GeorgSJPRB97,GeorgSETPRB98,GeorgSEMICRAS99}
covering the range of high temperatures and/or large conductance. 
It also is the basis of numerical calculations 
\cite{WangMCEPL97,ZwergerBOXPRL97,HerreroMCPRB99,GeorgSVDPhysicaB00}.

In this paper we study the single electron box by systematic 
diagrammatic techniques.
So far PT for any temperature has been calculated to the first
and the second orders in $\alpha$ \cite{GrabertBOXPRB94}. 
Here, we determine the third order corrections at zero temperature 
and compare them with Monte-Carlo (MC) results 
\cite{WangMCEPL97,ZwergerBOXPRL97,HerreroMCPRB99,GeorgBOXPRL98} 
and recent RG data \cite{KoenigRENPRB98}. 
Further, we explicitly address the vicinity of
the degeneracy point where PT fails and derive a
nonperturbative result by resummation of graphs contributing to 
the leading logarithmical divergencies.
We also discuss the charge noise
spectrum nonperturbatively by means of a projection operator 
technique. In general, the spectrum depends on the electronic bandwidth,
however, for frequencies relevant experimentally only the 
high frequency cutoff 
characterizing the resolution of the measuring device matters.

The paper is organized as follows:
In Sec.~\ref{sec:sysham} we
introduce the system Hamiltonian and the 
average island charge number.
In Sec.~\ref{sec:diagram} we briefly 
recapitulate essential results and diagrammatic rules of the
perturbative expansion \cite{GrabertBOXPRB94}. 
The necessary changes and simplifications
in the zero temperature 
limit are given in Sec.~\ref{sec:zeroT} where we 
exemplarily evaluate one graph of third order.
In Sec.~\ref{sec:chargenumber} we present the analytic result for
the average charge number. The two state approximation and the 
resummation of the leading logarithmical divergencies at the
degeneracy point are discussed
in Sec.~\ref{sec:resum}. In Sec.~\ref{sec:result} we compare
the analytical findings for the average 
island charge number and the effective charging energy with
MC data and recent RG results. Finally, in 
Sec.~\ref{sec:variance} we determine the noise spectrum of the 
island charge number and conclude in 
Sec.~\ref{sec:conclusions}.

\section{System and Model Hamiltonian} \label{sec:sysham}

\begin{figure}[btp]
\begin{center}
\leavevmode
\epsfxsize=0.3 \textwidth
\epsfbox{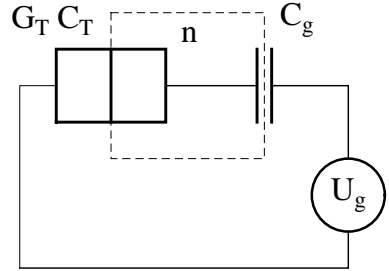}
\vspace*{-.0cm}
\caption{Circuit diagram of a single electron box.}
\label{fig:circuit}
\end{center}
\end{figure}

\noindent
We consider a SEB consisting of a metallic grain that couples to a lead
electrode via an oxide layer. 
The separation permits tunneling of single electrons with
the corresponding phenomenological tunneling conductance $G_T$. The 
geometrical capacitance between the grain an the lead reads $C_T$. 
Furthermore, a gate electrode is capacitively coupled to the grain
with gate capacitance $C_g$. 
This setup is shown schematically in Fig.~\ref{fig:circuit}, 
where the circuit is biased by a gate voltage
$U_g$ shifting the Coulomb energy continuously. We describe the 
SEB by the Hamiltonian \cite{GrabertBOXPRB94}

\begin{equation}
 H = H_0 + H_T
\end{equation}
where 
\begin{equation}
 H_0
 =
 H_C + H_{\rm qp}
\label{eq:hamil0}
\end{equation}
represents the system in absence of tunneling. Here,
\begin{equation}
 H_{\rm qp}
 =
 \sum_{k\sigma} 
     \epsilon_{k \sigma} a^\dagger_{k \sigma} a_{k \sigma} +
 \sum_{q\sigma} 
     \epsilon_{q \sigma} a^\dagger_{q \sigma} a_{q \sigma} 
\label{eq:hamilqp}
\end{equation}
describes free Fermions where the quasiparticle creation and
annihilation operators  
for transversal and spin quantum number 
$\sigma$ and longitudinal quantum number $p=k,q$ for the
island and the lead electrode, respectively, are denoted by 
$a^\dagger_{p \sigma}$ and $a_{p \sigma}$. Further, 
$\epsilon_{p \sigma}$ are quasiparticle energies 
for the corresponding quantum numbers.
\begin{equation}
 H_C
 = 
 E_C (n-n_g)^2
\label{eq:coulomben}
\end{equation}
is the Coulomb energy for $n$ excess charges on the island biased by
the dimensionless external voltage $n_g= C_g U_g/e$.
The charging energy 
\begin{equation}
 E_C
 = 
 \frac{e^2}{2C}
\end{equation}
depends solely on the island capacitance $C=C_T+C_g$. 
Spin and transversal quantum numbers are conserved during the tunneling 
process described by the tunneling Hamiltonian
\begin{equation}
 H_T
 =
 \sum_{kq\sigma} 
 \left(
   t_{kq\sigma} a^\dagger_{k\sigma} a_{q\sigma} \Lambda +\mbox{H.c.}
 \right),
\label{eq:tunnelHamil}
\end{equation}  
with $t_{kq\sigma}$ the transition amplitude between states with
quantum numbers $k \sigma$ and $q \sigma$. The
charge shift operator $\Lambda$ accounts for the Coulomb energy 
and is related to the charge number 
operator $n$ by the relation
\begin{equation}
 \Lambda^\dagger n \Lambda = n+1 .
\end{equation}
The excess charge number $n$ can be expressed as a derivative of the
system Hamiltonian with respect to $n_g$. Accordingly, we find 
for the average island charge number
\begin{equation}
 \langle n \rangle 
 = 
 n_g +
 \frac{1}{2 \beta E_c}~\frac{\partial \ln Z}{\partial n_g}~~,
\label{eq:nmittel1}
\end{equation}
where 
\begin{equation}
 Z
 = 
 {\rm tr}\{\exp(-\beta H)\}
\label{eq:ParitionGen}
\end{equation}
is the partition function of the system.
At low temperatures and
in the limit of vanishing tunneling conductance
the logarithm of the partition function reduces to the minimum of the
electrostatic energy $E_C(n_0-n_g)^2$ where $n_0$ is the integer
closest to $n_g$. Hence, as a function of the applied voltage
$U_g$ the island charge number displays the well known 
Coulomb staircase $\langle n \rangle = n_0$ observed
experimentally \cite{LafargeZPB91}. Due to occupation of higher energy
levels at finite temperatures the step function is smeared. Similarly,
the Coulomb staircase is rounded by virtual occupation of
higher charge levels caused by the finite tunneling conductance. 
Here, we
restrict ourselves to zero temperature and discuss the influence of 
higher order
tunneling processes on charge fluctuations.
Because of the periodicity and symmetry of the partition function 
$Z$ with respect to $n_g$, it is sufficient to consider
$0 \le n_g < \scriptstyle{\frac{1}{2}}$.

\section{Perturbation Expansion and Diagrammatic Rules}
 \label{sec:diagram}

\noindent
In this section we briefly summarize the method in 
Ref.~\cite{GrabertBOXPRB94} and give the diagrammatic rules used in
the remainder.

\subsection{Perturbation Expansion}
 \label{subsec:pertexp}

\noindent
Since the $n_g$ dependence of the partition function
$(\ref{eq:ParitionGen})$ arises
from the charging energy only, we may put
\begin{equation}
 Z
 :=
 \frac{{\rm tr} \, e^{-\beta H}}
    {{\rm tr}_{\rm qp} \, e^{-\beta H_{\rm qp}}} \,.
\label{eq:partfqp}
\end{equation}
Factorizing the exponential $\exp(-\beta H)$ 
into a part $\exp(-\beta H_0)$ in the absence of tunneling and
an interaction part written as a series in the
tunneling Hamiltonian $H_T$ we get
\begin{eqnarray}
 Z
&=&
 \sum_{m=0}^\infty (-1)^m
 \int_0^\beta d\alpha_m \int_0^{\alpha_m} d\alpha_{m-1} 
 \ldots \int_0^{\alpha_2} d\alpha_1 \, 
  \nonumber \\
&& 
 \frac{
   {\rm tr}
   \left\{ 
     e^{-\beta H_0} H_T(\alpha_m) \ldots H_T(\alpha_1)
   \right\}
      }{
   {\rm tr}_{\rm qp} \, e^{-\beta H_{\rm qp}}  
       } \, ,
\label{eq:boxIIpartfkt0}
\end{eqnarray}
with the tunneling Hamiltonian in imaginary time interaction
picture
\begin{equation}
 H_T(\tau)
 =
 e^{\tau H_0} H_T e^{-\tau H_0} \,.
\end{equation}
Separating the trace in a 
Coulomb and a quasiparticle trace 
\begin{equation}
 {\rm tr} \, e^{-\beta H_0} \ldots
 =
 {\rm tr}_C^{} \, e^{-\beta H_C} \, 
 {\rm tr}_{\rm qp} \, e^{-\beta H_{\rm qp}}
  \ldots
\end{equation}
one is left with multi-point 
correlation functions of tunneling Hamiltonians in 
imaginary time
\begin{eqnarray}
 Z
&=&
 \sum_{m=0}^\infty (-1)^m
 \int_0^\beta d\alpha_m \int_0^{\alpha_m} d\alpha_{m-1} 
 \ldots \int_0^{\alpha_2} d\alpha_1 \,
  \nonumber \\
&& 
 {\rm tr}_C^{} \, e^{-\beta H_C} \, 
   \langle 
     H_T(\alpha_m) \ldots H_T(\alpha_1)
   \rangle_0  \, ,
\label{eq:boxIIpartfkt1}
\end{eqnarray}
where
\begin{equation}
 \langle X \rangle_0
 =
 \frac{
   {\rm tr}_{\rm qp}
   \left\{ 
     e^{-\beta H_{\rm qp}}  \, X
   \right\}
      }{
   {\rm tr}_{\rm qp} \, e^{-\beta H_{\rm qp}}  
       } 
\end{equation}
is the thermal quasiparticle average.
Due to the Coulomb interaction  
these correlators do not decompose into a product of two-point 
correlators. However, inserting the explicit form 
$(\ref{eq:tunnelHamil})$ of $H_T$, 
the charge shift operators in interaction picture
$\Lambda(\tau)= \exp(\tau H_C) \Lambda \exp(-\tau H_C)$ 
commute with the quasiparticle operators and therefore
may be factored out of the quasiparticle trace leading to
\begin{eqnarray}
 Z
&=& 
 \sum_{m=0}^\infty (-1)^m
 \int_0^\beta d\alpha_m \int_0^{\alpha_m} d\alpha_{m-1} 
 \ldots \int_0^{\alpha_2} d\alpha_1     
\nonumber    \\
&&
 \sum_{k_1 q_1 \sigma_1 \zeta_1} \ldots
 \sum_{k_m q_m \sigma_m \zeta_m} 
 t^m \zeta_1 \ldots \zeta_m      
\nonumber    \\
&&
 \sum_{n=-\infty}^{\infty} 
    e^{-\sum_{j=1}^{2m} (\alpha_j-\alpha_{j-1}) E_{n_j}}     
     \label{eq:boxIIpartfkt}    \\
&&
 \left\langle 
   a_{k_m \sigma_m}^{\zeta_m}(\alpha_m)
   a_{q_m \sigma_m}^{-\zeta_m}(\alpha_m)
 \ldots
   a_{k_1 \sigma_1}^{\zeta_1}(\alpha_1)
   a_{q_1 \sigma_1}^{-\zeta_1}(\alpha_1)
 \right\rangle_0  \, . \nonumber 
\end{eqnarray}
Here, the Coulomb trace is explicitly represented as a sum over 
charge states labeled by $n$ and
\begin{equation}
 E_n
 = 
 E_C (n-n_g)^2  
\label{eq:coulomben1}
\end{equation}
is the corresponding Coulomb energy. The
charge shift operators in interaction picture lead to exponentials 
$\exp[\alpha_j ( E_{n_{j+1}} - E_{n_{j}})]$ where the integers
$n_j$ are defined by 
\begin{equation}
 n_1=n, \qquad \quad  n_j=\sum_{k=1}^{j-1}\zeta_k \, ,
\end{equation} 
with $\zeta_k=\pm$ labeling excess charge number increasing or 
decreasing processes.
Due to charge conservation the $\zeta$ sums in 
Eq.~$(\ref{eq:boxIIpartfkt})$ are constrained by
\begin{equation}
 \sum_{j=1}^{2m} \zeta_j =0  \, .
\label{eq:zetarestr}
\end{equation}
Further, we have introduced the shorthand notation 
$a^+=a^\dagger$ and $a^-=a$,
respectively, and $t=\overline{t_{kq\sigma}}$ is a real averaged
transmission amplitude.
Now time difference variables $\beta_j$, $(j=1, \ldots ,2m)$ 
between two subsequent operators are introduced and the cyclic
invariance of the trace can be used to obtain
\begin{eqnarray}
 Z
&=&
 \sum_{m=0}^\infty \frac{\beta}{2m}
     t^{2m}
 \int_0^\infty d\beta_1 \cdots \int_0^\infty d\beta_{2m}
 \delta \! \left(\sum_{j=1}^{2m}\beta_j-\beta \right)
         \nonumber  \\
& &
 \sum_{\zeta_1, \ldots , \zeta_{2m}} 
 \sum_{n=-\infty}^{\infty} e^{-\sum_{j=1}^{2m} \beta_j E_{n_j}}
 \sum_{k_1 q_1 \sigma_1} \cdots \sum_{k_{2m} q_{2m} \sigma_{2m}}
         \nonumber  \\
& &
 \left\langle 
   \prod_{j=1}^{2m} 
%
%
   \zeta_j a_{k_j \sigma_j}^{\zeta_j} \!
   \left( \sum_{l=1}^j \beta_l \right)
   a_{q_j \sigma_j}^{-\zeta_j} \!
   \left( \sum_{l=1}^j \beta_l \right)
 \right\rangle_0   \,  .
\label{eq:partition1}
\end{eqnarray}
Since the free quasiparticle Hamiltonian $H_{\rm qp}$
is quadratic in the fermionic degrees
of freedom, the expectations of quasiparticle operators 
$a^\dagger_{p\sigma}(\tau)$ and 
$a_{p\sigma}(\tau)$ obey a Wick theorem and the average in 
$(\ref{eq:partition1})$ 
decomposes into a sum over pair products of two-point 
correlators
\begin{equation}
 \left\langle
  a_{p_1\sigma_1}^{\zeta_1}(\tau_1) a_{p_2\sigma_2}^{\zeta_2}(\tau_2)
 \right\rangle_0
 =
 \delta_{\zeta_1,-\zeta_2} \delta_{p_1,p_2} 
 \delta_{\sigma_1, \sigma_2}
 \frac{e^{\zeta_1(\tau_1 -\tau_2)\epsilon_{p_1 \sigma_1}}}
      {1+e^{\zeta_1 \beta \epsilon_{p_1 \sigma_1}}}   .
\label{eq:towpointcor}
\end{equation}
So far the result is
valid for arbitrary numbers of tunneling channels
\begin{equation}
 {\cal N}
 =
 \sum_{\sigma} 1  .
\end{equation}
In metallic junctions where the junction area is typically 
much larger than the Fermi wavelength squared, ${\cal N}$ 
is very large. Experimentally the value is of the order 
${\cal N}\approx 10^4$ justifying limiting considerations. 
\begin{figure}[btp]
\begin{center}
\leavevmode
\epsfxsize=0.37 \textwidth
\epsfbox{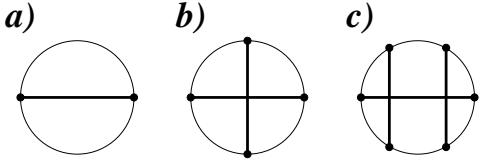}
\vspace*{-.0cm}
\caption{Representative circle diagrams of a) first, b) second, 
 and c) third order in the perturbative series.}
\label{fig:circlediag}
\end{center}
\end{figure}

\noindent
The $1/{\cal N}$ 
corrections for the SEB are considered explicitly in 
Ref.~\cite{GeorgChannelEPL99} confirming the validity of the
approximation. In leading order
only the combination 
\begin{eqnarray}
&& \!\!\!\!\!\! Y(\tau_2-\tau_1) 
           \nonumber     \\
&=&
 t^2 \sum_{k_1 q_1 k_2 q_2 \sigma}
 \langle 
   a^{\zeta}_{k_2 \sigma}(\tau_2) a^{-\zeta}_{k_1 \sigma}(\tau_1) 
 \rangle_0
 \langle 
   a^{-\zeta}_{q_2 \sigma}(\tau_2) a^{\zeta}_{q_1 \sigma}(\tau_1) 
 \rangle_0
           \nonumber     \\
&=&
 g \int_{-\infty}^{\infty} d\epsilon 
 \frac{\epsilon e^{-|\epsilon|/D}}{1-e^{-\beta\epsilon}}
e^{- (\tau_2-\tau_1) \epsilon}
\label{eq:greensf}
\end{eqnarray}
of two two-point correlators
contributes. Here, we have replaced the sums over $k$ and $q$ by 
energy integrals and have already performed one of them.
We introduced the notation
$g=t^2 {\cal N} \rho \rho' = \alpha/4\pi^2$ where $\rho$ and $\rho'$
are densities of states at the Fermi level for the island and the lead
electrode, respectively. 
The function $Y(\tau)$ is an electron-hole pair Green function
where electron and hole are created in different electrodes
and $D$ is the electronic bandwidth.
Representing the $\delta$-function in
Eq.~$(\ref{eq:partition1})$ in terms of an energy integral
over an auxiliary variable $E$
\begin{equation}
 \delta(\tau)
 = 
 \frac{1}{2 \pi} \int dE e^{-i\tau E}  \, ,
\end{equation}
one can perform all imaginary time integrals 
$\beta_j$, $(j=1 \ldots 2m)$ 
gaining energy denominators that are linear combinations of  
the auxiliary variable $E$ as well as 
$E_{n_j}$, $(j=1 \ldots 2m)$ and $\epsilon_k$, $(k=1 \ldots m)$. 
The coefficients of the
linear combinations depend on the Wick decomposition and
the $\zeta$-sums in $(\ref{eq:partition1})$. 
The remaining summations over pairs and $\zeta_j$'s can be
represented graphically by diagrams, in terms of which the 
partition function reads
\begin{equation}
 Z
 =
 \frac{1}{2 \pi} \int_{-\infty}^\infty dE \, e^{i \beta E}
 \sum_{\rm diagrams} D(E) \,.
\end{equation}
In Fig.~\ref{fig:circlediag} representative circle diagrams of a) 
first, b) second, and c) third order in PT are shown. 
A diagram $D(E)$ includes the charge 
sum $\sum_{n=-\infty}^\infty$ and each circle segment correspond 
to an energy denominator
\begin{equation}
 \frac{1}{E_{n_j} + \sum_{k=1}^m \Theta_{jk} \epsilon_k +iE} 
\label{eq:energydenom}
\end{equation} 
where $\Theta_{jk}=1$ for a time interval $\beta_j$ in between of two
vertices connected by a straight tunnelon line, and $\Theta_{jk}=0$ 
otherwise. The tunnelon lines represent energy integrals
\begin{equation}
 g \int d\epsilon_k 
 \frac{\epsilon_k e^{-|\epsilon_k|/D}}
      {1-e^{-\beta \epsilon_k}} 
\end{equation}
for $k=1, \ldots ,m$, stemming from 
electron-hole pair Green functions $(\ref{eq:greensf})$.
Since with respect to the auxiliary variable $E$
the whole integrand  
is a product of energy denominators,  
the integral can be performed explicitly by means of contour
integration. 
\begin{figure}[t]
\begin{center}
\leavevmode
\epsfxsize=0.2 \textwidth
\epsfbox{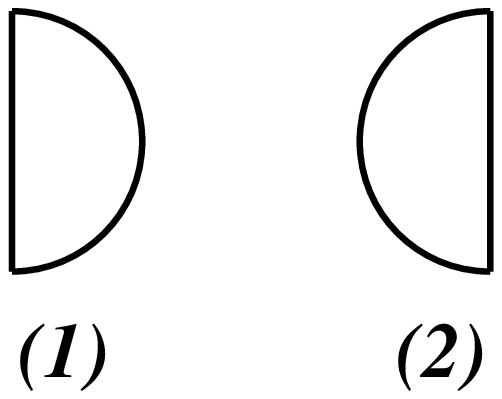}
\vspace*{-.0cm}
\caption{Diagrams of first order in the perturbative series.}
\label{fig:order1}
\end{center}
\end{figure}

\noindent
One gains a sum over residua corresponding to poles at a certain 
circle segment. Since residua of higher order poles correspond to
derivatives of the product of finite energy denominators
with respect to $E$, we get graphs decorated by slashes. 
The decorations are placed on the finite segments
indicating a derivative of the corresponding energy denominator 
with respect to the auxiliary variable $E$ evaluated at
the pole position considered. The simple form of the energy
denominators allows us to perform the derivatives explicitly. We gain
a higher power of the energy denominator $1/{\cal E}^{q+1}$ 
times $(-i)^q q!$ for a $q$-fold derivative of $1/{\cal E}$.
This way the sum over residua can be represented by circle diagrams, 
where the divergent circle segments are omitted and the remaining
finite segments are decorated with slashes indicating
derivatives of the energy denominator evaluated at the pole position.
Considering a circle diagram with
a pole of order $r$, one finds that when the pole segments are 
omitted, the graph decomposes into $r$ pieces 
denoted by $T_q$ with $q=1, \ldots ,r$. The partition function 
may then be arranged according to the order of poles
\begin{equation}
 Z
 =
 \sum_{r=1}^\infty  
 {\sum_{\rm diag}}^{r}  
 \sum_{n=-\infty}^\infty
 e^{-\beta E_n} \sum_{s=0}^{r-1} 
 \frac{\beta^{r-s}}{(r-s-1)!} \frac{1}{f}
 \left[ \prod_{q=1}^r T_q \right]^{(s)}
\end{equation}
where the sum over diagrams is restricted to those with poles of 
order $r$ and the symbol $[\ldots]^{(s)}$ stands for the sum over all 
decorations (derivatives) between the brackets with $s$ slashes.
Further, the factor $f$ is the number of identical 
subgroups in $T_q$, $q=1, \ldots ,r$.
Introducing the quantity 
\begin{equation}
 U_r^{(s)}
 =
 {\sum_{\rm diag}}^r \left[ \prod_{q=1}^r T_q \right]^{(s)}
\end{equation}
we may write the partition function in the form
\begin{equation}
 Z
 =
 \sum_{n=-\infty}^\infty e^{-\beta E_n}
 \left\{
  1+ \sum_{p=1}^\infty \frac{\beta^p}{p!}
     \sum_{s=0}^\infty \frac{p}{p+s} U_{p+s}^{(s)}
 \right\} \,.
\end{equation}
Using analytical properties of $U_r^{(s)}$ one may show that 
\cite{GrabertBOXPRB94}
\begin{equation}
 \sum_{s=0}^\infty \frac{p}{p+s} U_{p+s}^{(s)}
 = 
 \left(
  \sum_{s=0}^\infty \frac{1}{s+1} U_{s+1}^{(s)}
 \right)^p  \,.
\end{equation}
This way one gets an effective Coulomb representation 
of the partition function
\begin{equation}
 Z
 = 
 \sum_{n=-\infty}^{\infty}
 e^{-\beta(E_n+\Delta_n)} ,
\label{eq:coulombtrace}
\end{equation}
where 
\begin{equation}
 \Delta_n
 =
 - 
 \sum_{r=1}^\infty \,
 {\sum_{\rm diag}}^{r} \, \frac{1}{r}
 \left[ \prod_{q=1}^{r} \, T_q \right]^{(r-1)}
\end{equation}
is the energy correction to the Coulomb energy of
state $n$ that may be written
\begin{equation}
 \Delta_n
 = 
 \sum_{m=1}^\infty \Delta_n^{(m)} \, ,
\label{eq:energycorr}
\end{equation} 
where $\Delta_n^{(m)}$ is the contribution of order $g^m$.

\subsection{Diagrammatic Rules for $\Delta_n^{(m)}$}

\noindent
In this subsection 
we give the diagrammatic rules for the energy corrections
$(\ref{eq:energycorr})$. The term of order $g^m$ is given by
$\Delta_n^{(m)}$ and is composed of graphs containing a 
vertical line with $m$ semicircles attached. 
Each semicircle corresponds to a tunneling event and represents an
energy integral
\begin{equation}
 g \int_{-\infty}^\infty d\epsilon
 \frac{\epsilon e^{-|\epsilon|/D}}{1-e^{-\beta \epsilon}}
\label{eq:energintt}
\end{equation}
stemming from the electron-hole pair Green function 
$(\ref{eq:greensf})$.
Further each vertical line element contributes an energy 
denominator $\,-1/{\cal E}$ where ${\cal E}$ is the excitation energy 
during the corresponding intermediate state, which is the sum of the 
Coulomb energy difference 
\begin{equation}
 \xi_p = E_{n+p}-E_{n}
\label{eq:coulenergdiff}
\end{equation}
and of all electron-hole pair excitation energies 
$\epsilon_j$ present in the intermediate state that are represented  
by the arcs that would be intersected by a horizontal line. There 
are two types of semicircles: inflected to the right or left, 
whereby the Coulomb state $n$ is increased (decreased) 
by an arc to the right (left). 
For example in first order in $g$ there are just two
processes depicted in Fig.~\ref{fig:order1}. Whereas the graph $(1)$
increases the excess charge number by one and therefore
represents 
\begin{equation}
 \Delta_n^{(1,1)}
 =
 -g \int_{-\infty}^\infty d\epsilon
 \frac{\epsilon e^{-|\epsilon|/D}}{1-e^{-\beta \epsilon}}
 \frac{1}{\xi_1+\epsilon}   \, ,
\end{equation}
the graph $(2)$ lowers the charge number and it's contribution 
$\Delta_n^{(1,2)}$ is obtained 
from the previous one
by replacing $\xi_1 \rightarrow \xi_{-1}$.

\begin{figure}[btp]
\begin{center}
\leavevmode
\epsfxsize=0.3 \textwidth
\epsfbox{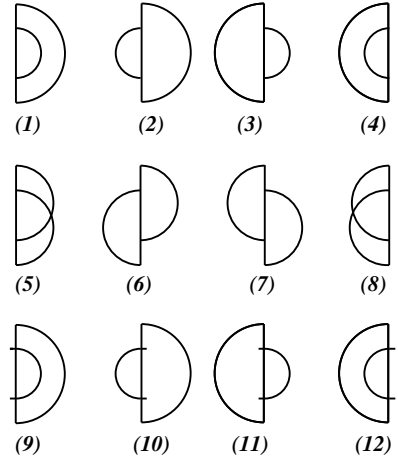}
\vspace*{-.0cm}
\caption{Diagrams of second order.}
\label{fig:order2}
\end{center}
\end{figure}

In second order, {\it cf}.\ Fig.~\ref{fig:order2}, there are 
two semicircles dividing the vertical
line into three parts. Each of them represents an energy denominator
at the corresponding charging energy.
The graphs $(1)$ to $(8)$ in Fig.~\ref{fig:order2} 
are given by all possibilities 
to attach two semicircles to a vertical line.
Using the rules given above the  
graph $(1)$ in Fig.~\ref{fig:order2} correspond to
\begin{eqnarray}
 \Delta_n^{(2,1)}
&=&
 -g^2 
 \int_{-\infty}^\infty d\epsilon_1   \int_{-\infty}^\infty d\epsilon_2
 \frac{\epsilon_1 e^{-|\epsilon_1|/D}}{1-e^{-\beta \epsilon_1}}
 \frac{\epsilon_2 e^{-|\epsilon_2|/D}}{1-e^{-\beta \epsilon_2}}
  \nonumber \\
&&
 \frac{1}{(\xi_1+\epsilon_1)^2}
 \frac{1}{\xi_2+\epsilon_1+\epsilon_2}
\end{eqnarray}
representing two tunneling processes with three intermediate energy
denominators. Since an 
arc to the left hand side represents a tunneling process  
that lowers the charge number on the island, 
the contribution of graph $(2)$ differs from that of graph 
$(1)$ by the replacement 
$\xi_2 \rightarrow 0$.
The first eight graphs in Fig.~\ref{fig:order2} can be generated
easily by the rules given. 
The graphs $(9)$ to $(12)$, however, differ by the 
prolongation of the
interior arcs across the vertical line. These ``insertions'' represent  
separate graphs, {\it i.e}.\ each insertion represents a graph of 
lower order multiplied to the main graph. However, the vertical
line of the main graph has two peaces with the same energy above and
below the insertion 
and therefore the energy denominator 
is squared. Moreover, each insertion carries a factor $(-1)$.

\begin{figure}[btp]
\begin{center}
\leavevmode
\epsfxsize=0.3 \textwidth
\epsfbox{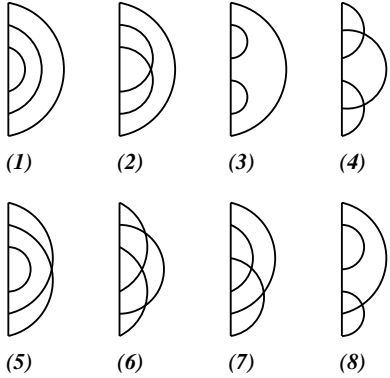}
\vspace*{-.0cm}
\caption{Representants of distinct graphs of third order without 
insertions.}
\label{fig:order3}
\end{center}
\end{figure}

\begin{figure}[btp]
\begin{center}
\leavevmode
\epsfxsize=0.3 \textwidth
\epsfbox{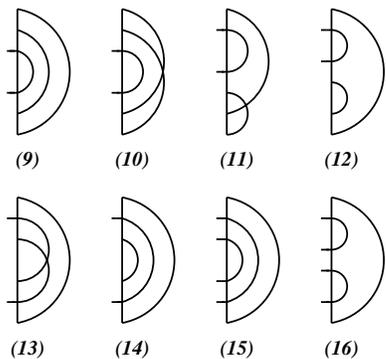}
\vspace*{-.0cm}
\caption{Representants of distinct graphs of third order with 
insertions.}
\label{fig:order3ins}
\end{center}
\end{figure}

\noindent
Analogously, with two insertions the main graph has a 
cubic energy denominator, see for example the graph $(16)$ 
in Fig.~\ref{fig:order3ins}. Using these rules, the
graph $(11)$ in Fig.~\ref{fig:order2} leads to the energy correction
\begin{eqnarray}
 \Delta_n^{(2,11)}
&=&
 g^2
 \int_{-\infty}^\infty d\epsilon_1 
 \int_{-\infty}^\infty d\epsilon_2
 \frac{\epsilon_1 e^{-|\epsilon_1|/D}}{1-e^{-\beta \epsilon_1}}
 \frac{\epsilon_2 e^{-|\epsilon_2|/D}}{1-e^{-\beta \epsilon_2}}
  \nonumber \\
&&
 \frac{1}{(\xi_{-1}+\epsilon_1)^2}
 \frac{1}{\xi_{1}+\epsilon_2}
\end{eqnarray}
that factorizes into $(-1)$ times the graph $(2)$ in 
Fig.~\ref{fig:order1} with the energy denominator squared, 
multiplied with 
the graph $(1)$ in Fig.~\ref{fig:order1}. 
These insertions are
shorthand notations for graphs with decorations that stem from a pole
of higher order in the energy denominator multiplied with a lower
order graph.
Therefore the higher order denominators and the factors $(-1)$
correspond to derivatives with respect to the auxiliary variable $E$.
At zero temperature these graphs are related to terms in the 
Rayleigh-Schr\"odinger perturbative expansion stemming from
normalization \cite{GeorgChannelEPL99}.

In the high temperature limit, or equivalently 
in the limit $E_C \rightarrow 0$, the Wick theorem implies that only 
``connected'' graphs appear in the exponent. The ``connected'' graphs are 
those where all processes occur in one channel.
In order $g^m$, $m>1$, these ``connected'' graphs,
however, give only $1/{\cal N}$ corrections that we have omitted. 
Hence, in the high temperature limit only the first order graphs 
survive and all higher order terms have to 
cancel. For the second order terms it turns out that each column in 
Fig.~\ref{fig:order2} cancels. For example, the contribution of the 
first column, graphs $(1),(5)$, and $(9)$, leads 
at vanishing charging energy to
\begin{eqnarray}
 && \hspace*{-.8cm} 
   \Delta_n^{(2;1,5,9)} \bigg|_{E_C \equiv 0}
        \nonumber  \\
&=&
 g^2
 \int_{-\infty}^\infty d\epsilon_1 
 \int_{-\infty}^\infty d\epsilon_2
 \frac{\epsilon_1 e^{-|\epsilon_1|/D}}{1-e^{-\beta \epsilon_1}}
 \frac{\epsilon_2 e^{-|\epsilon_2|/D}}{1-e^{-\beta \epsilon_2}}
        \nonumber  \\
& &    \times
 \left[
   \frac{1}{\epsilon_1^2(\epsilon_1+\epsilon_2)}  +
   \frac{1}{\epsilon_1(\epsilon_1+\epsilon_2)\epsilon_2} -
   \frac{1}{\epsilon_1^2 \epsilon_2}
 \right]
        \nonumber  \\
&\equiv&
 0  \, .
\end{eqnarray}
Higher order diagrams cancel likewise, and therefore in the 
limit of a large channel number the perturbative expansion can be
truncated after the first term at high temperatures where the 
charging energy $E_C$ is negligible compared to the thermal 
energy $k_B T$. 

The application of the diagrammatic rules to higher order graphs is 
obvious and we just present the contribution corresponding to 
graph $(7)$ in Fig.~\ref{fig:order3}
\begin{eqnarray}
 && \hspace*{-.4cm} 
 \Delta_n^{(3,7)}
                      \nonumber  \\
&=&  \!\!
 -g^3
 \int_{-\infty}^\infty d\epsilon_1 
 \int_{-\infty}^\infty d\epsilon_2
 \int_{-\infty}^\infty d\epsilon_3 
                      \nonumber  \\
& & \!\!
 \frac{\epsilon_1 e^{-|\epsilon_1|/D}}{1-e^{-\beta \epsilon_1}}
 \frac{\epsilon_2 e^{-|\epsilon_2|/D}}{1-e^{-\beta \epsilon_2}}
 \frac{\epsilon_3 e^{-|\epsilon_3|/D}}{1-e^{-\beta \epsilon_3}}
 \big[ (\xi_1+\epsilon_1)
      (\xi_2+\epsilon_1+\epsilon_2)
                      \nonumber  \\
& &  \!\!   \times
      (\xi_3+\epsilon_1+\epsilon_2+\epsilon_3)
      (\xi_2+\epsilon_1+\epsilon_3)(\xi_1+\epsilon_3)
 \big]^{-1}
\end{eqnarray}
as an example of a third order term.
Here, three tunneling processes occur and we have to deal with five
energy denominators. In this graph the excess charge number
is raised three times and then lowered stepwise. The energy
denominators include the quasiparticle excitation energies of arcs
that would be intersected by a horizontal line.
While the creation and annihilation times 
$\alpha_j$, $(j=1, \ldots ,2m)$ of these quasiparticle excitations 
are already integrated out, the order of creation and annihilation 
of different excitations is crucial. All types of diagrams of third 
order PT without insertions are depicted in
Fig.~\ref{fig:order3} and the graphs with one or two insertions 
are shown in Fig.~\ref{fig:order3ins}. Here, we have
omitted different combinations of semicircles to the left and
right but displayed only one representant. 
Hence, each graph in 
Figs.~\ref{fig:order3} and \ref{fig:order3ins} stands for
$8$ different left-right combinations 
of the $3$ semicircles. Further, the diagrams 
$(7)$ and $(8)$ in 
Fig.~\ref{fig:order3} and $(11)$ and
$(12)$ in Fig.~\ref{fig:order3ins} 
are topologically distinct from graphs reflected at a horizontal line,
but obviously, they lead to identical contributions and we 
have to count them twice. 

In summary, we have $80$ topologically 
different graphs without insertions, $64$ with one and $16$ graphs
with two insertions leading to $160$ topologically
different graphs of third order. All
contributions of third order are readily evaluated using the rules
given above.
In general, there are always $2^m$ left-right combinations for 
one representant of order $m$. Moreover, the number of representants 
exceeds the $(2m-1)!!$ possibilities to arrange $m$
semicircles along the vertical line, simply as a consequence of
the summation over insertions. 
Hence, the number of graphs of order $m$ exceeds $(2m)!/m!$,
{\it i.e}.\ grows faster than the factorial.
This rough estimation leads to more than $120$ graphs in third and 
more than $1680$ graphs in fourth order, showing that a reasonable
treatment is limited to third order. 
To obtain higher order or nonperturbative results, one has to
limit oneself to partial summations
of graphs including the essential contributions. 
Unfortunately, in the infinite cutoff limit $D \rightarrow \infty$,
each graph of the
perturbative series represents
a diverging integral and only the full sum in each order remains
finite \cite{GrabertBOXPRB94}. Hence,
partial summations of higher order graphs need an artificial cutoff 
complicating a direct
comparison with experimental findings. Consequently, within PT, 
the systematical treatment of higher orders is the
only tool to get results directly comparable with experiments. 
To proceed we discuss two general simplifications, 
valid for all orders.

\subsection{Reflected Graphs}

\noindent
The analytical form of the integrals leads to general 
consequences for energy corrections. First we note that 
the particular form of
the charging energy $(\ref{eq:coulomben})$ implies
\begin{equation}
 E_n(n_g) 
 = 
 E_{n+1}(n_g+1) 
 = 
 E_{-n}(-n_g)
\end{equation} 
and the corresponding energy differences in the denominators 
read
\begin{equation}
 \xi_{\pm p} 
 =
 E_{n \pm p}(n_g) - E_{n}(n_g)
 = 
 E_C [p^2 \pm 2p (n-n_g)]
\label{eq:coulombid}
\end{equation}
which depends, on the difference $(n-n_g)$ only.

\begin{figure}[btp]
\begin{center}
\leavevmode
\epsfxsize=0.3 \textwidth
\epsfbox{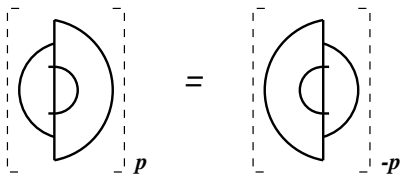}
\vspace*{-.0cm}
\caption{Symmetry of two vertically reflected graphs.}
\label{fig:nsymmetry}
\end{center}
\end{figure}

\noindent
Hence, the corrections of order $m$ may be written in the form
\begin{equation}
 \Delta_n^{(m)}(n_g)
 =
 g^m E_C f_m(n-n_g).
\end{equation}
Since contributions of
order $m$ include always $m$ integrals with $2m-1$ energy
denominators, we gain by measuring all energies in units of 
$E_C$ a single factor $E_C$, and
in the limit of an infinite bandwidth, $D\rightarrow \infty$, the
functions $f_m$ depend on $\beta E_C$ only. 
Further, a reflection of a given graph on the vertical axis
leads to the same contribution with the replacement
$E_{n\pm p} \rightarrow E_{n\mp p}$, 
{\it cf}.\ Fig.~\ref{fig:nsymmetry}. Equivalently, by virtue of 
Eq.~$(\ref{eq:coulombid})$, one can replace $(n-n_g)$ by $-(n-n_g)$.
Since the sum over diagrams includes all left-right 
permutations of arcs including pairs of vertically reflected
graphs, one may write
\begin{equation}
 f_m(u)=g_m(u)+g_m(-u) ,
\end{equation}
where the contribution of $g_m$ solely includes topologically different
graphs where one arc is held fixed.

\subsection{Insertions}  \label{subsec:insertions}

\begin{figure}[btp]
\begin{center}
\leavevmode
\epsfxsize=0.4 \textwidth
\epsfbox{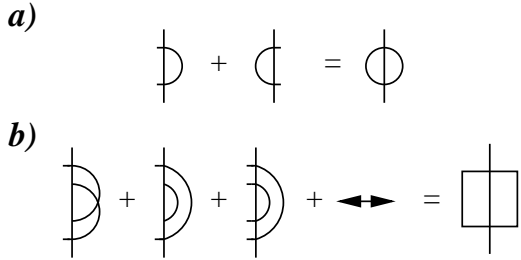}
\vspace*{-.0cm}
\caption{a) full first and b) full second order insertions.}
\label{fig:insertion}
\end{center}
\end{figure}

\noindent
A further simplification arises for graphs with insertions. It
turns out that they factorize into a host graph, with
energy denominator squared, and an
insertion contribution. Since we have to sum over all possible
left-right configurations, the full lower order 
contribution may be inserted. In
Fig.~\ref{fig:insertion} ${\rm a})$ we replace a first order 
insertion and its vertically reflected companion by a circle 
representing a
multiplication with the full first order contribution 
$\Delta_n^{(1)}$.
Likewise, insertions of order $k$ and all possible left-right 
combinations lead to a multiplication
with the full order $k$ contribution $\Delta_n^{(k)}$, 
schematically depicted for $k=2$ by 
the square in Fig.~\ref{fig:insertion} ${\rm b})$. The 
double arrow in this figure represents the
sum over all possible left-right arrangements of semicircles 
belonging to the insertion.
Therefore, all graphs belonging to the representants 
$(13),(14),$ and $(15)$ in
Fig.~\ref{fig:order3ins} lead to 
\begin{eqnarray}
 \Delta_n^{(3,13-15)}
&=&
 - g^3 E_C f_2(u) 
 \int_{-\infty}^\infty d\epsilon
 \frac{\epsilon e^{-|\epsilon| /D}}{1-e^{-\beta \epsilon}}
 \frac{1}{(\xi_{1} + \epsilon)^2} 
                      \nonumber  \\
& &    
   +   (u \rightarrow - u),
\end{eqnarray}
with $u=n-n_g$.
This is a multiplication of three factors, $\Delta_n^{(2)}$ 
and the host graph $(1)$ in Fig.~\ref{fig:order1} with the 
energy denominator squared, and $(-1)$. Additionally, 
we have added the contribution of the vertically
reflected graphs which is of the same form
with the replacement $u\rightarrow -u$.
Likewise, all graphs belonging to the representant $(16)$ in 
Fig.~\ref{fig:order3ins} lead to
\begin{eqnarray}
 \Delta_n^{(3,16)}
&=&
  -g^3 E_C^2 f_1(u)^2
 \int_{-\infty}^\infty d\epsilon
 \frac{\epsilon e^{-|\epsilon| /D}}{1-e^{-\beta \epsilon}}
  \frac{1}{(\xi_{1} + \epsilon)^3} 
                      \nonumber  \\
& &    
  +   (u \rightarrow - u)  \, ,
\end{eqnarray}
which is a multiplication of $\left( \Delta_n^{(1)} \right)^2$ 
and the host graph $(1)$ in Fig.~\ref{fig:order1} 
with the energy denominator cubed. This procedure holds for insertions
of all order and is used in Sec.~\ref{sec:resum} to resum the
leading logarithmic divergencies in the two-state approximation.

\section{Zero Temperature Limit and Third Order}
 \label{sec:zeroT}

\noindent
In this section we perform the zero temperature limit and 
present results of third order.

\subsection{Zero Temperature Limit}

\noindent
Generally, at zero temperature the partition function of a system with
nondegenerate discrete energy levels depends only on the ground 
state energy
\begin{equation}
 Z
 = 
 \sum_{n=-\infty}^{+\infty}
 e^{-\beta(E_n+\Delta_n)} 
 \longrightarrow
 e^{-\beta {\cal E}}  .
\end{equation}
Due to the restriction of the gate voltage to the range
$0 \le n_g < \scriptstyle{\frac{1}{2}}$, the ground state
charge number is $n=0$ and the ground state energy reads 
${\cal E}=E_0+\Delta_0$.
Therefore, the energy differences in the denominators read
\begin{equation}
 \xi_{\pm p} 
 = 
 E_{\pm p}-E_0
 =
 E_C(p^2 \mp 2p n_g)
\label{eq:energydiff}
\end{equation}
and the functions $f_m$ and $g_m$ depend on $n_g$ only.
The average island charge number reduces to
\begin{equation}
 \langle n \rangle 
 = 
 n_g -
 \frac{1}{2 E_c}~\frac{\partial {\cal E}}{\partial n_g}  \,.
\label{eq:nmittel2}
\end{equation}
Bose factors in the integrands restrict the integration 
$(\ref{eq:energintt})$ to positive energies
\begin{equation}
 g \int_{-\infty}^\infty d\epsilon
 \frac{\epsilon e^{-|\epsilon| /D}}{1-e^{-\beta \epsilon}}
 \longrightarrow 
 g \int_0^\infty d\epsilon
 \epsilon e^{-\epsilon /D}
\end{equation}
which facilitates the integration, because there are no
poles from the energy denominator 
$(\ref{eq:energydenom})$ 
that need to be taken care of, meaning there are no real excitations.

\subsection{First and Second Order Results}

\noindent
The first and second order calculations were already presented in
Ref.~\cite{GrabertBOXPRB94}.
The first order with an exponential cutoff $D$ leads to 
\begin{equation}
 g_1(u)
 =
 -(1+2u) \ln(1+2u) -D/E_C + \ln(D/E_C) + \gamma  , 
\label{eq:firstorder}
\end{equation}
where $\gamma=0.577 \ldots$ is Euler's constant. 
In the infinite cutoff limit this term diverges, but the
derivative with respect to $n_g$ remains finite. 
When we restrict ourselves to first order in $g$, we may omit 
this divergence,
however, in combination with higher order terms in the perturbative
series, {\it i.e.}\ as lower order insertion, we have 
to use the full expression $(\ref{eq:firstorder})$. 
Whereas the first order contribution diverges, the full sum of graphs 
of each higher order $m>1$ remains finite \cite{GrabertBOXPRB94}.
The second order contribution was also calculated previously 
leading to 
\begin{eqnarray}
& &  \hspace*{-0.4cm} 
 g_2(u) 
      \nonumber  \\
&=&
 \frac{\pi^2}{6}(1+2u+8u^2)
 -\left( \frac{5}{2}-4u+2u^2 \right) 
 \ln^2 \left( \frac{1-2u}{4(1-u)} \right)
      \nonumber  \\
& &
 + \left( \frac{1}{4}+u+u^2 \right)
 \ln^2 \left( \frac{1-2u}{1+2u} \right) 
      \nonumber  \\
& &
 +
 (1+2u) \ln \left( \frac{1-2u}{1+2u} \right) 
 - 4(1-u) \ln \left( \frac{1-2u}{4(1-u)} \right) 
        \nonumber \\
& &
 -
 (5-8u+4u^2) {\rm Li}_2 \left( \frac{3-2u}{4(1-u)} \right) ,
\end{eqnarray}
where ${\rm Li}_2(u)$ is the dilogarithm function \cite{Lewin81}.

\subsection{Third Order}

\noindent
Here, we motivate the calculation of the
third order contribution exemplarily for the graph 
$(7)$ in Fig.~\ref{fig:order3}. The full analytic expression
$g_3(u)$ is presented in the
Appendix.

Since the integral over the entire sum of $160$ integrands remains 
finite, we may omit the cutoff. However, to 
calculate the integral analytically 
we have to separate the whole expression into tractable parts. 
Each integral will be divergent and
we introduce a sharp high energy cutoff $D$. After integration  
we expand the expressions with respect to 
$D/E_C \rightarrow \infty$. There are 
divergent terms in each expression but the sum of divergencies of all
graphs has to vanish. This cancellation serves as a useful, 
nontrivial test of our 
calculation. We have to deal with eight different types of 
integrands without
insertions depicted in Fig.~\ref{fig:order3} and six 
different types with insertions in Fig.~\ref{fig:order3ins} 
(the graphs $(13)$ to $(15)$ are merged to a single graph with the
whole second order contribution inserted). 
We exemplarily proceed with
graph $(7)$ in Fig.~\ref{fig:order3} leading to an integral of 
type
\begin{eqnarray}
 h_7
&=& 
 \int_0^D d\epsilon_1 \int_0^D d\epsilon_2 \int_0^D d\epsilon_3
 \epsilon_1 \epsilon_2 \epsilon_3
   \nonumber \\
&&
  \big[ (\kappa_1+\epsilon_1) (\kappa_2+\epsilon_1+\epsilon_2) 
    (\kappa_3+\epsilon_1+\epsilon_2+\epsilon_3)
   \nonumber \\
&&  \times
    (\kappa_4+\epsilon_1+\epsilon_3) (\kappa_5+\epsilon_3)
  \big]^{-1} \,.
\label{eq:graphdritteord}
\end{eqnarray}
where the $\kappa_j$ stand for excitation energies of the
form $(\ref{eq:coulenergdiff})$.
The full contribution of this representant consists of all 
possible left-right arrangements of the semicircles leading to
\begin{equation}
 \Delta_0^{(3,7)}
 =
 -g^3 \sum  h_7(\kappa_1,\kappa_2,\kappa_3,\kappa_4,\kappa_5),
\end{equation}
where the sum runs over all allowed combinations of energy
differences. These are 
$h_7(\xi_1,\! \xi_2,\! \xi_3,\! \xi_2,\! \xi_1)$, 
$h_7(\xi_1,\! 0,\! \xi_1,\! \xi_2,\! \xi_1)$,
$h_7(\xi_1,\! 0,\! \xi_{-1},\! 0,\! \xi_{-1})$, 
$h_7(\xi_1,\! \xi_2,\! \xi_1,\! 0,\!  \xi_{-1})$, and
all terms with $n_g \rightarrow -n_g$.
The integrals may be performed by splitting denominators
into partial fractions. Using 
\begin{equation}
 \frac{1}{(a+\epsilon_2)(b+\epsilon_2)}
 =
 \frac{1}{(a-b)(b+\epsilon_2)}-\frac{1}{(a-b)(a+\epsilon_2)}
\label{eq:parfrac}
\end{equation}
where $a=\kappa_2+\epsilon_1$ and $b=\kappa_3+\epsilon_1+\epsilon_3$,
we are able to perform the $\epsilon_2$ integral leading to a logarithm
function
\begin{equation}
 \int_0^D d\epsilon_2 \frac{-\epsilon_2}{a+\epsilon_2}
 = 
 a \ln(a+D) -a \ln(a) -D.
\label{eq:logint}
\end{equation}
The new denominator $1/(a-b)$ on the rhs of 
Eq.~$(\ref{eq:parfrac})$ has an artificial pole at 
$\epsilon_3=\kappa_2 - \kappa_3$. Since the integrand is
analytic in the integration region, the sum of all pole contributions 
has to cancel in the threefold integral $(\ref{eq:graphdritteord})$. 
However, each pole contribution depends on the contour of
integration and we have to specify the contour and use the same
for all integrals. 
Next we consider the $\epsilon_1$ integration.
In the numerator now appears the logarithm function 
from the previous integration,
$\ln(\kappa_2+\epsilon_1)$, that diverges for
$\epsilon_1\rightarrow 0$ when $\kappa_2\equiv 0$ and we 
temporarily introduce a lower integration limit.
With a decomposition of the form $(\ref{eq:parfrac})$, where 
$\epsilon_2$ is replaced by $\epsilon_1$ and 
the constants read $a=\kappa_1$ and $b= \kappa_4+\epsilon_3$, we
split the remaining fractions in $(\ref{eq:graphdritteord})$
and perform the integral in terms of the 
dilogarithm function \cite{Lewin81}
\begin{equation}
 {\rm Li}_2(z)
 =
 -\int_0^z dz' \frac{\ln(1-z')}{z'} .
\end{equation} 
The third integration can then be performed using the trilogarithm 
function ${\rm Li}_3(z)$, where the
general polylogarithm functions are defined by \cite{Lewin81}
\begin{equation}
 {\rm Li}_n(z)
 =
 \int_0^z dz' \frac{{\rm Li}_{n-1}(z')}{z'} .
\end{equation} 
All terms emerging can be expressed by trilogarithms,
dilogarithms, logarithms, and rational functions of $n_g$.
Here, the arguments of the transcendental functions are rational
expressions of energy differences. Since each integration 
increases the ``order'' of transcendental functions at most 
by one, the transcendental terms are of the form
${\rm Li}_k \ln^l$ obeying $k+l \le 3$. Here $\ln^l$ stands for a 
product of $l$ logarithms of possibly different arguments. 
The ``order'' thereby characterizes the transcendental function for
large arguments: {\it e.g}.\ ${\rm Li}_k \sim \ln^k$ is of order $k$. 
We find in order $m$ of the perturbative series 
transcendental terms of the form 
$({\rm Li}_{m})^{k_m} \cdots ({\rm Li}_{2})^{k_2} \ln^{k_1}$ 
where $\sum_{j=1}^m  j k_j \le m$. 
In principle, the integrals in PT lead in all orders 
to analytically known functions but there are practical
restrictions, in particular, since the integrals cannot be 
evaluated straightforwardly by tools like {\it Mathematica} 
because one has to take care of the integration contours and 
pole contributions explicitly. In view of the length
of the analytical result we present $g_3(u)$ in the
Appendix. 

The result is in terms of complicated analytical
functions that have to be calculated numerically. 
Therefore, one could think of a direct numerical evaluation 
of the three fold integrals, but there are
serious numerical problems. First, only the full sum of the $160$
integrals is finite and one has to use a huge integrand. 
Moreover, the integrand is not symmetric in the three integration
variables, in particular, there are integration directions 
where the integrand leads to diverging positive and negative 
contributions. For a numerical study one has to symmetrize 
the integrand which enlarges it by a factor of $6$. 
Second, in spite of the smoothness and analyticity of the integrand 
and the absence of poles in the integration region, the integrand 
contains oscillatory parts so that standard numerical routines 
fail. We used a statistical integration routine to check the 
analytical predictions for selected parameter values. Typically, 
the numerical evaluation of a single point with a few per cent 
accuracy took over a week of CPU on a {\it SGI Origin 200}. 
Therefore, a numerical evaluation of the integral in third order 
is hopeless, in particular for delicate high precision studies 
regarding the limiting behavior near the degeneracy point or 
calculations of second order derivatives needed to determine 
the charging energy.

\section{Average Charge Number}  \label{sec:chargenumber}

\noindent
The analytical result for the average island charge 
number $\langle n \rangle$ at zero temperature and in 
first order in $g$ was calculated previously
\cite{MatveevBOXJETP91,EsteveNato92in}
\begin{equation}
 \langle n \rangle_1^{}
 =
 g\ln\frac{1+2n_g}{1-2n_g} 
\label{eq:avchargefirst}
\end{equation}
and can be readily obtained by using
Eq.~$(\ref{eq:nmittel2})$ with the energy correction 
$(\ref{eq:firstorder})$.
The function is well behaved except at the degeneracy point  
$n_g = \scriptstyle{\frac{1}{2}}$ where it
exhibits a logarithmic divergence 
$\langle n \rangle_1^{} \sim -g\ln\delta$, with 
$\delta = {\scriptstyle \frac{1}{2}}-n_g$.
Therefore, the range of
validity of the perturbative series at zero temperature is 
strictly restricted 
to $n_g < \scriptstyle{\frac{1}{2}}$.
The contribution of second order in
$g$ presented in Ref.~\cite{GrabertBOXPhysica94} reads
\begin{eqnarray}
&&  \hspace*{-0.4cm}
  \langle n \rangle_2 
  \nonumber \\
&=&
   -g^2\Bigg\{n_g\left[\frac{4\pi^2}{3}+
   \ln^2\left(\frac{1-2n_g}{1+2n_g}\right)\right] 
  \\
&&   
   +
   \frac{16(1+2n_g-2n_g^2)}{(3-2n_g)(1+2n_g)} 
   \ln(1-2n_g) 
\label{eq:nsecondord} 
  \nonumber \\
&& 
   +
   2(1-n_g)\Bigg[ 
   \ln^2
    \left(
      \frac{1-2n_g}{4(1-n_g)}
    \right)+ 2
   \mbox{Li}_2 
   \left(
      \frac{3-2n_g}{4(1-n_g)}
   \right) 
  \nonumber \\
&& 
   -\frac{8(1-n_g)}{(1-2n_g)(3-2n_g)}
   \ln(4(1-n_g)) \Bigg] 
   - (n_g \rightarrow  - n_g) \Bigg\} \,.
  \nonumber
\end{eqnarray}
Here, $(n_g \rightarrow  - n_g)$ stands for the same sum of 
terms with $n_g$ replaced by
$-n_g$ showing explicitly the asymmetry of 
$\langle n \rangle$ with respect to the applied voltage $U_g$.
At the degeneracy point, a leading logarithmic divergency of 
$\langle n \rangle _2 \sim - 2 g^2 \ln^2 \delta$ 
appears. This indicates that near the degeneracy point $g\ln\delta$ 
is the effective expansion parameter so that the larger $g$ the 
smaller is the range of $n_g$ where finite order PT suffices. 
The analytical result of third order is not given explicitly here 
but can easily be calculated by differentiating the expression 
$\Delta_0^{(3)}$ with respect to $n_g$, 
{\it cf}.\ $(\ref{eq:nmittel2})$. Also the third order term 
shows a logarithmic divergency at the degeneracy point leading 
together with the lower order contributions to the asymptotic 
expansion 
\begin{eqnarray}
\langle n \rangle &=&
 ag^2+bg^3-(g+6g^2+cg^3)\ln\delta  
  \nonumber \\
&&  
  -(2g^2+24g^3)\ln^2\delta
  -4g^3\ln^3\delta + {\cal O} \left(\delta, g^4 \right) \,,  
\label{eq:nmittel3}
\end{eqnarray}
where the coefficients $a$ and $c$ read
\begin{eqnarray}
 a 
&=&
 3 - \frac{5}{3}\pi^2 - \frac{9}{2}\ln(3) 
 + 3\ln^2(3) + 6 \,{\rm Li}_2 \left( \frac{2}{3} \right) 
   \nonumber  \\  
&=&  -9.7726 \ldots  \, ,
\label{eq:nmittel31}  \\
 c
&=&
 21 + \frac{21}{4}\pi^2 + 9\ln(3) 
 - 6\ln^2(3) -12 \,{\rm Li}_2 \left( \frac{2}{3} \right)
   \nonumber  \\  
&=& 65.462 \ldots  \, .
\label{eq:nmittel32}
\end{eqnarray}
In view of its length the coefficient $b$ is given only
numerically, $b=-70.546...$, but it can be readily obtained 
from the analytical result of $\Delta_0^{(3)}$ in the Appendix.
The leading order logarithmic terms in
Eq.\ (\ref{eq:nmittel3}) read
\begin{equation}
\langle n \rangle \sim
-g\ln\delta-2g^2\ln^2\delta-4g^3\ln^3\delta  \,.
\label{eq:nmittel4}
\end{equation} 
They are related to diagrams that contain only the charge states
$n=0$ and $1$. Therefore, before comparing our findings with 
numerical results, we consider a two-state
approximation restricted to these degenerate charge states.

\section{Degeneracy Point} \label{sec:resum}

\begin{figure}[btp]
\begin{center}
\leavevmode
\epsfxsize=0.14 \textwidth
\epsfbox{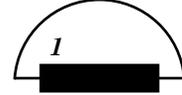}
\vspace*{-.0cm}
\caption{Full ground state energy diagram in the non-crossing  
 approximation of the two-state model.}
\label{fig:ground}
\end{center}
\end{figure}
\begin{figure}[btp]
\begin{center}
\leavevmode
\epsfxsize=0.43 \textwidth
\epsfbox{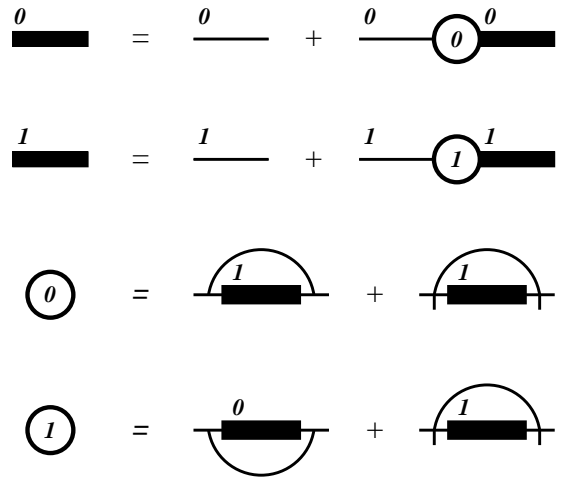}
\vspace*{-.0cm}
\caption{Dyson equation in the non-crossing approximation
  of the two-state model.}
\label{fig:dyson}
\end{center}
\end{figure}

\noindent
A further analysis of the perturbative series in the limit of 
$n_g \rightarrow \scriptstyle{\frac{1}{2}}$ shows
that the leading logarithmic divergencies $(\ref{eq:nmittel4})$
stem from
diagrams including only charge states $n=0$ and $1$. This is a direct
consequence of the degeneracy of these charge states at
$n_g=\scriptstyle{\frac{1}{2}}$. The two-state approximation 
limits the perturbative series  
to graphs that contain the charge states $n=0,1$ only. 
This model was shown to be equivalent to an anisotropic 
multi channel Kondo Hamiltonian where $E_C \delta$ corresponds to 
the magnetic field and $g$ to the exchange integral
\cite{MatveevBOXJETP91}.
Considering
the leading order logarithmic divergencies we find that crossed diagrams,
like graph $(6)$ in Fig.~\ref{fig:order3} with the middle semicircle
to the left, do not contribute. Therefore, we may restrict ourselves 
to noncrossing diagrams. It is then possible to write the 
ground state energy as
\begin{equation}
 \widetilde{\Delta}_0
 =
 g \int_0^D d\epsilon \epsilon  F_1(\epsilon)  
\end{equation}
graphically represented in Fig.~\ref{fig:ground}.
The generalized energy denominator $F_1(\epsilon)$ 
corresponds to the bold line with index $1$ and is determined by 
a Dyson equation graphically represented in Fig.~\ref{fig:dyson}.
For convenience, we have rotated the graphs by $90$ degrees and an
upper (lower) semicircle increases (lowers) the charge number. 
Further, since we sum a subset of graphs, we need to introduce 
a cutoff $D$, which for convenience is chosen as a sharp cutoff.
In the Dyson equation depicted in Fig.~\ref{fig:dyson}
the thin line with index $0$ corresponds to the bare energy
denominator $-1/\epsilon$ of the charge state $0$ where 
$\epsilon$ is the energy variable. Analogously, the thin line
with index $1$ correspond to the bare energy denominator 
$-1/(\xi_1 + \epsilon)$ of charge state $1$. On the other hand, 
the bold line with index $0$ $(1)$ represents the
dressed $n=0$ $(n=1)$ propagator 
$F_0(\epsilon)$ $(F_1(\epsilon))$ arising from the insertion of 
interaction vertices according to the first (second) line 
in Fig.~\ref{fig:dyson}. 
Hereby the interaction vertex represented by a circle with 
index $0$ $(1)$ is composed
of a semicircle and an insertion according to the third 
(fourth) line in Fig.~\ref{fig:dyson}. These diagrams are 
evaluated by employing the rules given in 
Sec.~\ref{sec:diagram}. Each part of an interaction vertex 
contains an integral 
$g\int_0^D d\epsilon' \epsilon'$ but the two pieces differ 
in the explicit meaning of $\epsilon'$: While the insertion 
is just a multiplication 
where the propagator inside does not depend on the energy variable 
$\epsilon$ of the legs of the vertex, the
semicircle propagator takes into account all excitation energies 
and therefore depends on $\epsilon+\epsilon'$.
Due to the restriction of the perturbative series to the 
charge states $n=0$ and $n=1$, the interaction vertices 
contain only two parts shown in Fig.~\ref{fig:dyson}, and
the Dyson equation generates all diagrams in the
non-crossing approximation. However, to be consistent
with the truncation to two charge states, one has to introduce
a cutoff $D$ that restricts electron-hole pair excitations 
to energies lower than the next charge state. With higher 
energy excitations being eliminated the parameters must be 
interpreted as effective renormalized quantities. 
\begin{figure}[btp]
\begin{center}
\leavevmode
\epsfxsize=0.45 \textwidth
\epsfbox{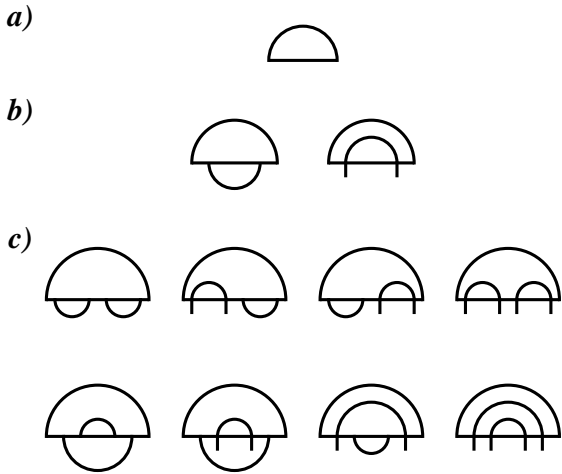}
\vspace*{-.0cm}
\caption{Graphs of a) first, b) second, and c) third
order in the two-state model generated by the 
Dyson equation in Fig.~\protect\ref{fig:dyson}.}
\label{fig:2state}
\end{center}
\end{figure}
\noindent
One way to find the renormalization of the parameters is a
comparison of the limiting divergent behavior for 
$n_g \rightarrow \scriptstyle{\frac{1}{2}}$ with the full PT
result. This comparison leads to a series expansion of the 
renormalized parameters $g^*$ and $E_C^*$ in terms of the 
bare conductance $g$. 

\begin{figure}[btp]
\begin{center}
\leavevmode
\epsfxsize=0.25 \textwidth
\epsfbox{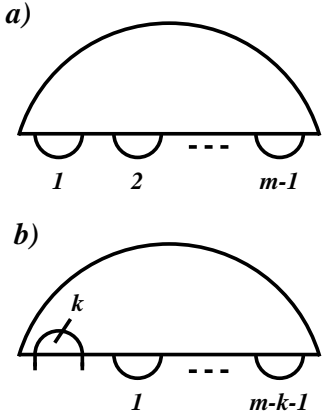}
\vspace*{-.0cm}
\caption{Diagrams of order $g^m$ contributing to the leading
logarithmic divergencies.}
\label{fig:leading}
\end{center}
\end{figure}

Iterating the Dyson equation we generate graphs including 
only charge states $n=0,1$. 
In Fig.~\ref{fig:2state} we depict all graphs of 
a) first, b) second, and c) third order in $g$ obtained 
this way. However, not all of
these graphs contribute to the leading behavior near
$n_g \rightarrow \scriptstyle{\frac{1}{2}}$. We find that 
only graphs of the form shown in Fig.~\ref{fig:leading} are
responsible for the leading logarithmic divergencies in order
$g^m$. The diagram ${\rm a})$ shows $m-1$ downward arcs
leading to the asymptotic behavior
\begin{equation}
 \widetilde{\Delta}_0^{(m,0)} 
 \sim
 - g^m \delta \ln^m \delta 
\end{equation}
for $\delta \rightarrow 0$. The diagram ${\rm b})$ shows 
$m-k-1$ arcs lowering the charge state and one multi-insertion 
represented as an insertion with a slash with label 
$k$ $(k<m)$ which stands for the full result in the two-state
approximation in order $k$. Assuming that the order $k$ result
has the leading order behavior
\begin{equation}
 \widetilde{\Delta}_0^{(k)} 
 \sim
 -\frac{1}{2} \left(2 g \ln \delta \right)^k \delta
\label{eq:assumption}
\end{equation}
as is apparent for $k=1,2$ and $3$ form 
Eq.~$(\ref{eq:nmittel4})$, these graphs 
asymptotically behave as 
\begin{equation}
 \widetilde{\Delta}_0^{(m,k)} 
 \sim
 \widetilde{\Delta}_0^{(k)} \, 
 g^{m-k} \frac{\ln^{m-k}\delta}{m-k}
 =
 - g^m \, \frac{2^{k-1}}{m-k} \,  \delta \ln^m \delta \, .
\label{eq:dysonkinsert}
\end{equation}
Interchange of the multi-insertion and the arcs leads to
topologically different contributions that have to be counted 
separately. This leads to a factor $(m-k)$ canceling the denominator
in Eq.~$(\ref{eq:dysonkinsert})$.
The sum over all $k$'s starting with $k=0$ as represented by graph 
${\rm a})$ in Fig.~\ref{fig:leading} and from $k=1$ to $k=m-1$ 
corresponding to graph ${\rm b})$ leads to an asymptotic behavior
of the form  
\begin{equation}
 \widetilde{\Delta}_0^{(m)} 
 \sim
 \sum_{k=0}^{m-1} \widetilde{\Delta}_0^{(m,k)} 
 =
 -\frac{1}{2} \left(2 g \ln \delta \right)^m \delta 
\end{equation}
which proves the assumption $(\ref{eq:assumption})$ for all $k$ 
by induction.
For the average island charge number $(\ref{eq:nmittel2})$ 
we obtain by summing over all orders in $g$
\begin{equation}
\langle n \rangle =
\frac{-g\ln\delta}{1-2g\ln\delta}  \,.
\label{eq:matveev}
\end{equation} 
This result was 
originally obtained by Matveev
\cite{MatveevBOXJETP91} with RG techniques. While this result 
does not depend explicitly on the cutoff $D$, it contains
renormalized parameters $g$ and $\delta$ depending on the tunneling 
strength. The leading logarithmic divergencies up to third order 
coincide with those given in Eq.~$(\ref{eq:nmittel4})$, however, 
nonleading divergencies from Eq.~$(\ref{eq:nmittel3})$ are missing. 
They can be included by using renormalized parameters 
$g^*$ and $\delta^*$ in Eq.~$(\ref{eq:matveev})$. 
Since the energy difference between the two charge states is
$E_C \delta$, the renormalization of the charging energy may be
written as $E_C \delta^*$. The renormalized quantities are
then found to read
\begin{equation}
 g^* 
 = 
 g \left[ 1 + 6g + c_2 g^2 + {\cal O}(g^3) \right]
\label{eq:grenormal}
\end{equation}
and 
\begin{equation}
 \delta^*
 =
 \delta \left[ 1 + d_1 g + d_2 g^2 + {\cal O}(g^3) \right] 
\label{eq:drenormal}
\end{equation}
where the renormalization factors are given as series in powers 
of $g$ with coefficients that may be expressed in terms of 
the expansion coefficients in 
Eqs.~$(\ref{eq:nmittel3})$--$(\ref{eq:nmittel32})$. We have  
$c_2=-4a-c=-26.372 \ldots$, $d_1=-a=9.7726 \ldots$, 
and $d_2=6a+a^2/2-b=59.662 \ldots$.
As a consequence of the rapidly increasing
series coefficients the result $(\ref{eq:matveev})$ 
with $(\ref{eq:grenormal})$ and $(\ref{eq:drenormal})$ 
does not lead to meaningful
results in the vicinity of $n_g=\scriptstyle{\frac{1}{2}}$
except for very small $g$.
Therefore, one has to resum also nonleading divergencies in the
two-state approximation to describe the behavior near 
$n_g=\scriptstyle{\frac{1}{2}}$. This
has not been considered so far but is the aim of future work. 

\section{Discussion of Results} 
               \label{sec:result}

\begin{figure}[ht]
\begin{center}
\leavevmode
\epsfxsize=0.48 \textwidth
\epsfbox{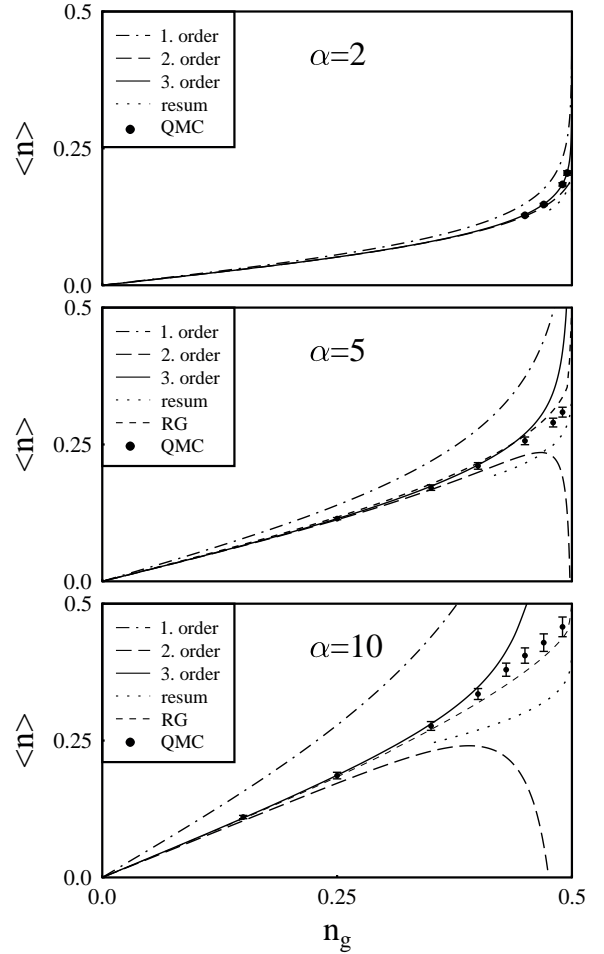}
\vspace*{-.0cm}
\caption{The average electron number $\langle n \rangle$ as a
function of the dimensionless voltage $n_g$ is shown in first, second,
and third order perturbation theory in $\alpha$, and compared with QMC
data \protect\cite{GeorgBOXPRL98} and RG results 
\protect\cite{KoenigRGPRL98}. 
The result $(\protect\ref{eq:matveev})$ is also shown 
as a dotted line.}
\label{fig:thirdstep}
\end{center}
\end{figure}

\noindent
In this section we compare our analytical results with numerical data
and estimate the range of validity of various orders of PT. 
While various QMC studies of the single electron box are available
\cite{WangMCEPL97,ZwergerBOXPRL97,HerreroMCPRB99,GeorgBOXPRL98}, only the
data in Ref.~\cite{GeorgBOXPRL98} determine the island 
charge number for finite gate voltages and very low temperatures so
that they can be compared with zero temperature predictions. 
Further, we confront our results with the findings of a recent 
renormalization group study \cite{KoenigRGPRL98}.
The comparison of first, second, and
third order PT with QMC data \cite{GeorgBOXPRL98} and 
RG results \cite{KoenigRGPRL98} for $\alpha=2, 5$, 
and $10$ is shown in Fig.~\ref{fig:thirdstep}. 
Not to overload the graph for $\alpha=2$ the RG results are 
omitted but they coincide with third order PT.
We give results here in terms of the dimensionless tunneling 
conductance $\alpha=G_T/G_K=4\pi^2 g$. 
We find good agreement for gate
voltages near zero, but for 
$n_g \rightarrow \scriptstyle{\frac{1}{2}}$
the analytic result diverges. As discussed above the range of 
validity of PT shrinks with increasing
gate voltage. 
We find that third order PT in
$\alpha$ remains valid with errors
below $4\%$ up to $n_g \approx 0.495$ for dimensionless conductance
$\alpha=2$, up to $n_g \approx 0.45$ for $\alpha=5$, and
up to $n_g \approx 0.4$ for $\alpha=10$. In these parameter intervals
PT agrees both with QMC and RG data. 
Since for $n_g=0.45$
the charging energies for $n=0$ and
$n=1$ differ only by $0.1 E_C$, deviations from the third order result
in $\alpha$ can be observed only for temperatures well below
$E_C/10k_B$ even at $n_g={\scriptstyle \frac{1}{2}}$.
Hence, at finite temperature the range of validity of PT 
increases. Further, Fig.~\ref{fig:thirdstep} shows that 
the resummation of the leading
logarithmic terms $(\ref{eq:matveev})$ does not suffice to describe
the behavior
near $n_g = {\scriptstyle \frac{1}{2}}$. Subleading logarithms
are important to obtain quantitatively meaningful results in the
strong tunneling regime. We remark that the inclusion of 
subleading logarithmic terms of low orders only, in 
terms of the renormalized parameters $g^*$ and $\delta^*$
defined in Eqs.~$(\ref{eq:grenormal})$ and $(\ref{eq:drenormal})$, 
does not improve the agreement, rather one has to consistently 
resum the nonleading logarithmic terms to all orders.

Since for $n_g \rightarrow 0$ the QMC and RG data perfectly 
coincide with
the perturbative results, we used a more sensitive quantity to 
determine the range of validity of PT in this limit:
For small external voltages, the average island charge grows linearly as
\begin{equation}
\langle Q \rangle = e\langle n \rangle = C^* U_g
\end{equation}
where $C^*$ is an effective capacitance of the box. In the absence of
Coulomb blockade effects $C^*=C$, while for strong Coulomb
blockade, i.e., in the limit of vanishing tunneling conductance,
$C^*=0$. It is thus natural to characterize the strength of the Coulomb
blockade effect by an effective charging energy $E_C^*$ defined by
\cite{WangMCEPL97}
\begin{equation}
 \frac{E_C^*}{E_C} 
 =
 1-\frac{C^*}{C}
 =
 1-
 \left. 
   \frac{\partial \langle n \rangle }{\partial n_g} 
 \right|_{n_g=0} .
\label{eq:is0}
\end{equation}
One may view $E_C^*$ as the effective rounding of the energy
parabolas at $n_g=0$ in the presence of tunneling.
The perturbative series gives
\begin{equation}
\frac{E_C^*}{E_C} =
 1-4g+Ag^2-Bg^3+{\cal O}(g^4),
\label{eq:u}
\end{equation}
where 
\begin{eqnarray}
 A
&=&
 \frac{8\pi^2 - 32}{3} + \frac{64}{9}\ln(2)
 - 16\ln^2(2) - 8\, {\rm Li}_2 \left(\frac{3}{4}\right)
    \nonumber  \\
&=&
 5.066...
\end{eqnarray}
and $B=1.457...$ is given here only numerically in view of the 
length of the analytical expression.

\begin{figure}[btp]
\begin{center}
\leavevmode
\epsfxsize=0.5 \textwidth
\epsfbox{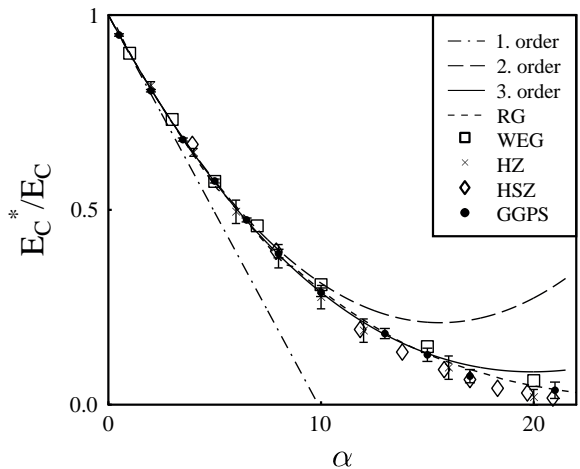}
\vspace*{-.0cm}
\caption{Effective charging energy as a function of the dimensionless
conduction $\alpha=G_T/G_K$. Perturbative findings are compared with
QMC data  
\protect\cite{WangMCEPL97,ZwergerBOXPRL97,HerreroMCPRB99,GeorgBOXPRL98}
and RG results 
\protect\cite{KoenigRGPRL98}.}
\label{fig:thirdec}
\end{center}
\end{figure}

In Fig.~\ref{fig:thirdec} we compare our predictions for the 
effective charging energy with QMC data by 
Wang, Egger, and Grabert \cite{WangMCEPL97} (WEG), 
Hofstetter and Zwerger \cite{ZwergerBOXPRL97} (HZ), 
Herrero, Sch\"on, and Zaikin \cite{HerreroMCPRB99} (HSZ), 
and G\"oppert {\it et al.} \cite{GeorgBOXPRL98} (GGPS)
and RG results by K\"onig and Schoeller \cite{KoenigRGPRL98}.
We find good agreement with PT up to $\alpha \approx 8$ 
for second order PT, while third order PT extends to 
$\alpha \approx 16$. 
Discrepancies between these QMC studies only arise outside
of the range of validity of third order PT.

\section{Fluctuations of the average charge number} 
   \label{sec:variance}

In this section we study the steady state fluctuations
of the average charge number $\langle n \rangle$.
We are mainly interested in the variance 
$\langle \delta n^2 \rangle$ with
$\delta n = n - \langle n \rangle$. Since 
$(n-n_g)^2=\partial H_C/\partial E_C$, one may calculate this 
quantity in terms of a derivative of the free energy. 
At zero temperature, the free energy to first 
order PT is given by $E_C [g_1(n_g)+g_1(-n_g)]$ with
the function $g_1$ defined in Eq.~$(\ref{eq:firstorder})$. 
However, this result would lead to a logarithmic divergence 
of $\langle \delta n^2 \rangle$ for large electronic
bandwidth $D$. The same divergence occurs also at finite 
temperatures,
{\it cf.} Ref.~\cite{GrabertBOXPRB94}. 
To explore where this cutoff dependence comes
from, we consider the correlation function 
$\langle \delta n(t) \delta n(0) \rangle$. Using PT
this quantity is well defined for finite $t$ but diverges 
in the limit $t\rightarrow 0$. This divergence is found to 
result from a $1/\omega$ tail for large $\omega$ in the 
fluctuation spectrum arising from the coupling to high 
energy electron-hole excitations. While this tail is cut 
off by the finite bandwidth $D$, in real experiments 
there is a cutoff from the time resolution of the 
measurement device at a certain frequency $\omega^{}_M$ well 
below $D/\hbar$. To avoid these cutoff effects, we discuss 
only the noise spectrum
\begin{equation}
 S_n(\omega)
 =
 \int dt e^{i \omega t} C(t)
\end{equation}
that is the Fourier 
transform of the symmetrized correlation function
\begin{equation}
 C(t)
 =
 \frac{1}{2} \langle \delta n(t) \delta n(0) 
 + \delta n(0) \delta n(t) \rangle
 \equiv \langle \delta n(t) \delta n(0) \rangle^s \, .
\end{equation}
While $S_n(\omega)$ also depends on the electronic bandwidth, 
for frequencies well below the electronic cutoff and not too low 
temperatures the spectrum becomes independent of the 
bandwidth. Here, 
$\delta n(t) = \exp(i{\rm L}t)n - \langle n \rangle$ is 
the fluctuation of the average charge number in the 
Heisenberg representation, where the
time evolution is governed by
the Liouville operator ${\rm L}X= \frac{1}{\hbar} [H,X]$.

Since a direct perturbative expansion of $C(t)$ results
effectively in a short time expansion, it is insufficient 
to determine the low frequency behavior of $S_n(\omega)$ 
properly. Here, we use projection operator techniques 
\cite{GrabertPOT82} to derive a formally exact integral 
equation for the dynamics of $C(t)$ and then expand the 
kernel in powers of the
dimensionless conductance $g$. The adequate projector reads
\begin{equation}
 PX 
 = 
 \frac{\delta n}
      {\langle \delta n^2 \rangle} 
      \langle \delta n X \rangle^s
\end{equation}
which fulfills the requirement $P^2=P$. Applying the 
time evolution operator one finds
the relation
\begin{equation}
 P e^{i{\rm L}t} P
 = 
 \bar{C}(t) P \, ,
\label{eq:projcid}
\end{equation}
with the normalized correlator 
$\bar{C}(t)=C(t)/\langle \delta n^2 \rangle$. Further, the time
derivative of the correlator may be written in the form
\begin{equation}
 \dot{\bar{C}}(t)P
 =
 P e^{i{\rm L}t}i{\rm L}P.
\label{eq:correlderiv}
\end{equation}
Using the operator identity
\begin{equation}
 Pe^{i{\rm L}t}
 =
 Pe^{i{\rm L}t}P + 
 \int_0^t ds P e^{i{\rm L}s} P i {\rm L} \bar{P} 
 e^{i\bar{P}{\rm L} \bar{P}(t-s)} \, ,
\end{equation}
with $\bar{P}=1-P$, Eq.~$(\ref{eq:correlderiv})$ 
may be written as
\begin{eqnarray}
 \dot{\bar{C}}(t) P
&=&
 P e^{i{\rm L}t} P i {\rm L} P 
 \nonumber   \\
&&
 + \int_0^t ds P e^{i {\rm L}s}P i{\rm L} \bar{P}
 e^{i\bar{P}{\rm L} \bar{P}(t-s)} i{\rm L} P \, .
\end{eqnarray}
Inserting the identity $(\ref{eq:projcid})$ we get the
exact evolution equation
\begin{equation}
 \dot{C}(t)
 =
 \frac{\langle \delta n \dot{n} \rangle^s}
      {\langle \delta n^2 \rangle}
 C(t) 
 -
 \int_0^t ds \phi(t-s) C(s) \, ,
\label{eq:cintegeq}
\end{equation}
with the memory kernel
\begin{equation}
 \phi(t)
 =
 \frac{\langle \dot{n}\dot{n}_r(t) \rangle^s}
      {\langle \delta n^2 \rangle}
 -
 \frac{\langle \delta n \dot{n} \rangle^s}
      {\langle \delta n^2 \rangle}
 \frac{\langle \delta n \dot{n}_r(t) \rangle^s}
      {\langle \delta n^2 \rangle}    \, .
\end{equation}
Here, the reduced time evolution of $\dot{n}$ is given by 
\begin{equation}
 \dot{n}_r(t)
 =
 e^{i\bar{P}{\rm L} \bar{P} t} \dot{n} \, .
\end{equation}
The linear equation $(\ref{eq:cintegeq})$ can be solved by 
means of a Laplace transformation yielding
\begin{equation}
 \widehat{C}(z)
 = 
 \frac{C(0)}{z-\langle \delta n \dot{n} \rangle^s/
      \langle \delta n^2 \rangle + \widehat{\phi}(z)} 
\label{eq:noiselapl}
\end{equation}
which is related to the noise spectrum by
\begin{eqnarray}
 S_n(\omega) \!
&=&  \!
 \int_{-\infty}^\infty dt e^{i\omega t} C(t)
 =
 2 {\rm Re} \int_0^\infty dt e^{i\omega t} C(t)
       \nonumber  \\
&=& \!
 2 {\rm Re} \! \lim_{z\rightarrow -i \omega} \! 
 \int_0^\infty \!\!\! dt e^{-z t} C(t)
 \!=\! 
   2 {\rm Re} \lim_{z\rightarrow -i \omega} 
   \widehat{C}(z) \, .
\label{eq:noiselaplfour} 
\end{eqnarray}
To evaluate the right hand side of 
Eq.~$(\ref{eq:noiselapl})$ we first determine the 
initial value 
$C(0)=\langle n^2 \rangle - \langle n \rangle^2$ 
in second order in the tunneling Hamiltonian
by means of a straightforward expansion of the 
density matrix using the imaginary time methods 
in Sec.~\ref{subsec:pertexp}.
Since $n$ and $n^2$ are diagonal in charge representation,
both averages may be written in the form
\begin{eqnarray}
 && \!\!\!\!\!\! \langle X \rangle 
 =
 \frac{1}{Z}{\rm tr}
 \left\{
    e^{-\beta H} X
 \right\}
 \nonumber               \\
&=&
 \frac{1}{Z}{\rm tr}
 \Big\{
   e^{-\beta H_0}
   \Big[
     1-\int_0^\beta \!\! d\tau' H_T(\tau')
     + 
 \nonumber               \\
&&   \quad
      \int_0^\beta d\tau' \int_0^{\tau'} \!\! d\tau'' 
        H_T(\tau') H_T(\tau'')
   \Big]
   X
 \Big\}
\label{eq:imagexp}
               \\
&=&
 \frac{Z_0}{Z}
 \Big\langle 
   X+ \!\sum_{\zeta=\pm}
   \int_0^\beta \!\!\! d\tau' \int_0^{\tau'} \!\!\! d\tau'' 
   Y(\tau'-\tau'') e^{-(\tau'-\tau'')\xi_\zeta}
  X
 \Big\rangle_C \, ,
\nonumber
\end{eqnarray}
with the free partition function $Z_0={\rm tr}\exp(-\beta H_0)$ 
and the Coulomb energy differences $\xi_\zeta$  that were 
introduced in Eq.~$(\ref{eq:coulenergdiff})$.
Here, we have decomposed the trace into a Coulomb and a 
quasiparticle trace as in Sec.~\ref{subsec:pertexp}, and 
have introduced the free Coulomb average
\begin{equation}
 \langle X \rangle_C^{} 
 =
 \sum_{n=-\infty}^{\infty} e^{-\beta E_n} X \Bigg/ 
 \sum_{n=-\infty}^{\infty} e^{-\beta E_n}  \, .
\end{equation}
Inserting the representation $(\ref{eq:greensf})$ 
for the electron-hole pair propagator into
Eq.~$(\ref{eq:imagexp})$,
the time integrals are readily evaluated leading to an 
energy integral that may be solved by contour integration,
{\it cf.} Ref.~\cite{GrabertBOXPRB94}. For the average 
charge number squared we find
\begin{eqnarray}
 \langle n^2 \rangle
&=&
 \langle n^2 \rangle_C (1+ \langle K_1 \rangle_C)
    \nonumber \\
& &
 - \langle n^2 K_1 
 + 2 n K_2 
 + K_3 \rangle_C  
 + {\cal O}(g^2) \, .
\end{eqnarray}
The first term $\langle K_1 \rangle_C$ stems from the
expansion of the denominator $Z$ where
\begin{equation}
 K_1
 = 
 2\pi g {\rm Im} 
 \sum_{\zeta=\pm} [f_1(iE_C/\nu)- f_1(i\xi_\zeta/\nu)] 
\end{equation}
and $\nu=2 \pi/ \beta$. A divergent part for 
$D \rightarrow \infty$ is omitted since it is 
canceled by a corresponding term in the expansion
of the numerator.
Further, we introduced the auxiliary function
\begin{equation}
 f_1(x)=x \psi(1+x)  
\end{equation}
where $\psi(z)$ is the logarithmic derivative of the 
gamma function. The remaining terms contain
\begin{equation}
 K_2
 = 
 g
 \sum_{\zeta=\pm} \zeta {\rm Re}[f_1'(i\xi_\zeta/\nu)]
\label{eq:variancek2}
\end{equation}
and 
\begin{equation}
 K_3
 =
 g
 \left\{
   2 [\gamma - \ln(D/\nu)]
   +
   \sum_{\zeta=\pm} {\rm Re}[f_1'(i\xi_\zeta/\nu)]
 \right\}               \, .
\label{eq:K3withco}
\end{equation}
Whereas $K_1$ and $K_2$ are independent of the cutoff, the 
last function diverges in the infinite bandwidth limit.
For the average charge number we get likewise
\begin{equation}
 \langle n \rangle
 =
 \langle n \rangle_C (1+ \langle K_1 \rangle_C)
 - \langle n K_1 +  K_2 \rangle_C  \, .
\end{equation} 

Next, we determine
$\langle \delta n \dot{n} \rangle^s$ where
the time derivative of the charge number
\begin{equation}
 \dot{n} 
 =
 \frac{i}{\hbar} [H,n]
 = 
 \frac{i}{\hbar} \sum_{kq\sigma} 
 \left(
   t_{kq\sigma} a^\dagger_{k\sigma} a_{q\sigma} \Lambda
   - {\rm H.c.}
 \right)
\end{equation}
arises from the tunneling Hamiltonian.
Since $\dot{n}$ is of first order in the tunneling Hamiltonian 
and is off-diagonal in 
charge representation, we have to expand the density matrix up to 
first order, 
{\it cf.} Eq.~$(\ref{eq:imagexp})$, and
one readily finds 
$\langle \delta n \dot{n} \rangle^s = {\cal O}(g^2)$.
Now, the only term contributing in second order in the
tunneling Hamiltonian reads
\begin{eqnarray}
 \langle \dot{n} \dot{n}_r(t) \rangle^s
&=&
 \frac{1}{Z_0} {\rm Re}
 \left[
 {\rm tr} 
 \left\{
   e^{-\beta H_0} \dot{n} 
   e^{\frac{i}{\hbar} H_0t} \dot{n} e^{-\frac{i}{\hbar} H_0t}
 \right\} 
 \right] 
      \nonumber  \\
&=& 
 g
 {\rm Re}
 \left\langle 
   \sum_{\zeta=\pm} \int d\epsilon
   \frac{\epsilon e^{-|\epsilon|/D}}{1-e^{-\beta \epsilon}}
   e^{-\frac{i}{\hbar}(\epsilon +\xi_\zeta)t}
 \right\rangle_C \, ,
\end{eqnarray}
where we have used that
$\exp(i\bar{P}L_0 \bar{P}t)\dot{n}
 =\exp(\frac{i}{\hbar} H_0 t)\dot{n}\exp(-\frac{i}{\hbar} H_0 t)
  +{\cal O}(g)$.
The Laplace transform can be readily performed leading to
\begin{eqnarray}
 \! \widehat{\phi}(z) \!
&=&
 \frac{g}{2 \langle \delta n^2 \rangle_C}
 \sum_{\zeta,\zeta'=\pm} 
 \Bigg\langle 
   \frac{\pi}{\hbar}
   \frac{-\xi_\zeta + i\zeta' \hbar z}
        {1-e^{\beta(\xi_\zeta -i\zeta' \hbar z)}}
   e^{-| \xi_\zeta + i\zeta' \hbar z |/D}
       \nonumber \\
&&  \hspace*{-1.0cm}
   + z \! \sum_{m=1}^\infty  \!
     \frac{m e^{-|\nu_m|/D}}
     {[m \! - (i\zeta' \xi_\zeta-\hbar z)/\nu] 
      [m-(i\zeta' \xi_\zeta+\hbar z)/\nu]} \!
 \Bigg\rangle_C  \hspace*{-0.3cm}
\label{eq:kernelD}
\end{eqnarray}
which depends on the electronic bandwidth $D$ and 
shows a $1/z$ dependence for $z\gg D/\hbar$. The  
spectral density of the charge fluctuations
now follows from Eqs.~$(\ref{eq:noiselapl})$ and 
$(\ref{eq:noiselaplfour})$.
Since the measurement 
device suppresses frequencies above a cutoff  
$\omega^{}_M \ll D/\hbar$, we may focus on 
$|z| \ll D/\hbar$ and find 
\begin{eqnarray}
 \widehat{\phi}(z)
&=&
 \frac{g}{2 \langle \delta n^2 \rangle_C}
 \sum_{\zeta,\zeta'=\pm} 
 \Bigg\langle 
   \frac{\pi}{\hbar}
   \frac{-\xi_\zeta + i\zeta' \hbar z}
        {1-e^{\beta(\xi_\zeta -i\zeta' \hbar z)}} 
       \nonumber \\
&&   \quad
   - \zeta'
   \frac{\nu}{2 \hbar} \sum_{\kappa=\pm}
     f_1\left( 
          \frac{i \kappa \xi_\zeta +\zeta' \hbar z}{\nu} 
        \right)
 \Bigg\rangle_C
       \nonumber \\
&&   \quad
  + 
 2gz \frac{\ln(D/\nu)-\gamma}{\langle \delta n^2 \rangle_C} 
\label{eq:kernel1zllD}
\end{eqnarray}
which determines the noise spectrum for frequencies 
$|\omega| \ll D/\hbar$:
\begin{equation}
 S_n(\omega)
 =
 2 \mbox{Re} 
 \frac{C(0)}{-i\omega + \widehat{\phi}(-i\omega)}
  \, .
\label{eq:noisespec}
\end{equation}
While $\widehat{\phi}(z)$ has been evaluated to 
first order in $g$ the spectrum $S_n(\omega)$ 
contains terms of all orders in $g$.

Two approximations can be considered:

\noindent
{\it i}) For large 
$|\omega| \gg \exp(-\langle \delta n^2 \rangle_C/g)D/\hbar$
we may expand the denominator in Eq.~$(\ref{eq:noisespec})$
\begin{equation}
 \frac{C(0)}{-i\omega + \widehat{\phi}(-i\omega)}
 =
 C(0)
 \left[
  \frac{1}{-i\omega} 
  + 
  \frac{\widehat{\phi}(-i\omega)}{\omega^2}
  + {\cal O}(g^2)
 \right] 
\end{equation}
and get the perturbative result which reads for
$|\omega| \ll D/\hbar$
\begin{eqnarray}
 S_n(\omega)
&=&
  2\pi C^f(0) \delta(\omega)
  \nonumber   \\
&& \hspace*{-0.8cm}
 +
 g \frac{\pi}{\hbar \omega^2}  \sum_{\zeta \zeta'=\pm}
 \left\langle
  \frac{\zeta' \hbar \omega-\xi_\zeta}
   {1-e^{-\beta(\zeta' \hbar \omega-\xi_\zeta)}}
 \right\rangle_C 
 + {\cal O}(g^2)  \, ,
\label{eq:noisepert} 
\end{eqnarray}
in accordance with earlier findings
\cite{SchoenPRB85}. 
Here, the diverging parts of $C(0)$ in Eq.~$(\ref{eq:K3withco})$ 
and of $\widehat{\phi}(-i\omega)$ in 
Eq.~$(\ref{eq:kernel1zllD})$ cancel leading to the finite
contribution $C^f(0)=C(0)+\langle K_3 \rangle_C$
independent of the cutoff $D$.
However, this approximation  
is not valid at small frequencies and
shows a $1/\omega^2$ divergence that cannot describe the 
low frequency behavior correctly. In contrast, for large 
frequencies the result $(\ref{eq:noisepert})$ is well behaved 
and merges with our result $(\ref{eq:noisespec})$. 

\noindent
{\it ii})
On the other hand, for small 
$z  \ll  \pi k_B T/\gamma \hbar$ at large temperatures
$k_B T \gg E_C$  and/or
$z  \ll  \pi E_C \exp(-\beta E_C)/\hbar \ln(\beta E_C)$
at moderate to low temperatures 
$k_B T 
{\tiny{\lower 2pt \hbox{$<$}\atop\raise 5pt \hbox{$\sim$}}}
E_C$ 
we may replace $\widehat{\phi}(z)$ by 
$\widehat{\phi}(0) + \widehat{\phi}'(0)z$
which is often 
referred to as Markovian approximation. Then
\begin{equation}
 \widehat{C}(z)
 =
 \frac{C(0)}{z[1+\widehat{\phi}'(0)]+\widehat{\phi}(0)}
\end{equation} 
where 
$\widehat{\phi}'(0) 
 = 
 - \langle K_3 \rangle_C/ \langle \delta n^2 \rangle_C $.
In the limit 
$2 g \ln(D/\nu) \ll \langle \delta n^2 \rangle_C$ we may 
expand $1+\phi'(0)$ around $g=0$ and get
\begin{equation}
 S_n(\omega)
 =
 C^f(0)
 \frac{2 \widehat{\phi}(0)}{\omega^2+\widehat{\phi}(0)^2} \, .
\label{eq:markovomega}
\end{equation}
While the Markovian approximation fails to describe the high 
frequency behavior, at high temperatures the classical result 
\begin{equation}
 \widetilde{C}_M(\omega)
 =
 \frac{k_B T C}{e^2} 
 \frac{2 G_T/C}{\omega^2 + (G_T/C)^2}
\end{equation}
is recovered. Hence, at high temperatures the 
first order approximation $(\ref{eq:kernel1zllD})$
of the memory kernel $\phi(t)$ becomes
exact.

Now, provided the experimental cutoff $\omega^{}_M$ is small 
enough for the Markovian approximation to be justified we have
\begin{equation}
 \langle \delta n(t) \delta n(0) \rangle_M^s
 =
 \frac{1}{2\pi}
 \int d\omega S_n(\omega) F(\omega/\omega^{}_M) e^{-i\omega t}
\end{equation}
with a cutoff function $F(\omega/\omega^{}_M)$ obeying 
$F(0)=1$. Using contour integration we get
in the limit $t\rightarrow 0$
\begin{equation}
 \langle \delta n^2 \rangle_M
 =
 C^f(0) F(\widehat{\phi}(0)/\omega^{}_M)
 =
 C^f(0) + {\cal O}( \widehat{\phi}(0)/\omega^{}_M) \, .
\label{eq:varianceMarkov}
\end{equation}
Hence, for $\widehat{\phi}(0) \ll \omega^{}_M \ll D/\hbar$
the measured variance is
independent of the cutoff.

\begin{figure}[btp]
\begin{center}
\leavevmode
\epsfxsize=0.48 \textwidth
\epsfbox{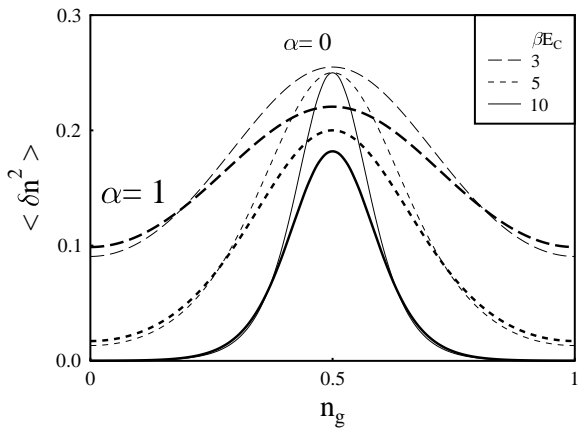}
\vspace*{-.0cm}
\caption{Variance of the average charge number in dependence
on the dimensionless gate voltage for different temperatures. 
Depicted are the results for $\alpha=0$ (thin lines) compared 
with results for $\alpha=1$ (bold lines).}
\label{fig:noise}
\end{center}
\end{figure}

To show the impact of finite tunneling conductance we compare
in Fig.~\ref{fig:noise} the variance 
$(\ref{eq:varianceMarkov})$ of the average 
charge number in the Markovian approximation for $\alpha=0$
(thin lines) and $\alpha=1$ (bold lines) for different 
temperatures in dependence on 
the dimensionless gate voltage $n_g$. Whereas tunneling 
amplifies the noise near $n_g=0$, in the vicinity of the
degeneracy point the noise is suppressed by tunneling leading
to the asymptotic behavior 
$\langle \delta n^2 \rangle_M \approx {\scriptstyle \frac{1}{4}}-
 g [\ln( \beta E_C/\pi) +1+\gamma]$.
Since fluctuations of the charge are related to the linear
conductance of the single electron transistor (SET) this 
behavior is in accordance with the observation that
the linear conductance of the SET increases with the 
tunneling strength off the degeneracy point and on the other 
hand decreases directly at the degeneracy
point, {\it cf.} 
Refs.~\cite{GeorgSETPRB98,JoyezSETPRL97,KoenigCOTPRL97}.

Finally, we remark that directly at the degeneracy point
$\widehat{\phi}(z)$ becomes nonanalytic at zero temperatures
for $z\rightarrow 0$ and the Markovian
approximation breaks down.
Therefore, at lower temperatures $k_BT \ll E_C$ one has to use the
full bandwidth dependent expression $(\ref{eq:kernelD})$.

\section{Conclusions} \label{sec:conclusions}
In this article charge fluctuations of the single electron box 
were investigated by means of perturbation theory. Terms of 
third order in the phenomenological tunneling conductance were 
calculated analytically
in the zero temperature limit. The predictions for the average charge
number and the effective charging energy have been compared with
Monte-Carlo data and renormalization group results. It has been shown
that the perturbative treatment, in spite of its diverging behavior at
the degeneracy point, leads to reliable results in a large part
of the parameter regime explored experimentally. An additional 
order in the perturbative series increases the range of validity
considerably, in particular for small gate voltages 
$n_g \approx 0$ where PT covers the largest range of conductance
parameters $\alpha$. When increasing the gate voltage the range 
shrinks continuously down to zero at the degeneracy point 
$n_g={\scriptstyle \frac{1}{2}}$.  
In the vicinity of the degeneracy point we have evaluated all
graphs contributing to the leading 
logarithmic divergencies. A resummation was found to be 
insufficient to describe the behavior
of the average charge number in the strong tunneling
regime quantitatively. It turns out that nonleading logarithmic
divergencies are essential, the summation of which remains an open
problem. 

Further, the noise spectrum of the average charge number
has been investigated. In contrast to the average charge number
this quantity depends on the electronic bandwidth $D$. 
Using a projection operator
technique we have obtained an expression covering the perturbative 
as well as the semiclassical 
results. We find that in the Coulomb blockade region 
near $n_g=0$ strong
tunneling leads to an increase of the noise whereas at the 
degeneracy point $n_g={\scriptstyle \frac{1}{2}}$ 
the noise is suppressed.

\section*{Acknowledgments}

The authors would like to thank M.~H.\ Devo\-ret, D.\ Esteve, 
P.\ Joyez, J.\ K\"onig, N.~V.\ Prokof'ev, H.\ Schoeller, and
B.~V.\ Svistunov for valuable discussions.
Financial support was provided by the Deutsche
Forschungsgemeinschaft (DFG) and the Deutscher Akademischer 
Austauschdienst (DAAD).

\end{multicols}                                   %
\widetext                                         %

\begin{appendix}

\section*{Third Order Result} 
\label{app:thirdres}

\noindent
In this appendix we give the explicit analytic result of
the contribution in the third order in $g$ to the ground 
state energy ${\cal E}$. According to the transcendental 
functions appearing in the various terms
we split the result into six contributions
\begin{equation}
 g_3(u)
 = 
 P3(u) + P2(u) + L3(u) + L2(u) + L1(u) + R(u) \,.
\end{equation}
Here $P3$ contains all terms with trilogarithms ${\rm Li}_3$ 
while $P2$ lists those with dilogarithms ${\rm Li}_2$. 
Terms containing logarithms and no other
transcendental functions are 
split into three types: In $L3$ and $L2$ expressions 
containing $\ln^3$ and $\ln^2$ are listed, respectively. 
Simple logarithms appear in 
$L1$ and the remaining rational functions of 
$u$ and constants are gathered in $R$.
With the abbreviation 
\begin{equation}
 \xi_n^m= \xi_m-\xi_n
\end{equation}
these terms read

\begin{eqnarray}
&&\!\!\!\!\!\!\!\!\!\!   P3(u) =  \\
&&\!\!\!\!\!\!\!\!\!\!
 \left( 
  {\frac{207}{4}} + {\frac{223\,u}{2}} + 105\,{u^2} + 34\,{u^3} \
 \right) 
 \,{\rm Li}_3 \left({\frac{\xi_1}{\xi_2}}\right) 
 + 
 \left( 
  {\frac{79}{2}} + {\frac{211\,u}{2}} + 102\,{u^2} + 34\,{u^3} \
 \right) 
 \,{\rm Li}_3\left({\frac{\xi_1^2}{\xi_2}}\right)
         \nonumber \\
&&\!\!\!\!\!\!\!\!\!\!
+ \xi_1\,
  \left[ 
    {\rm Li}_3\left({\frac{\xi_{-1}^2}{2\,\xi_2}}\right) - 
     2\,{\rm Li}_3\left({\frac{\xi_1}{2\,\xi_1^2}}\right) + 
     {\rm Li}_3\left({\frac{\xi_1}{2\,{\xi_1^3}}}\right) + 
     {\rm Li}_3\left({\frac{\xi_2^3}{2\,\xi_2}}\right) - 
     2\,{\rm Li}_3\left({\frac{\xi_2^3}{2\,\xi_1^2}}\right) + 
     {\rm Li}_3\left({\frac{3\,\xi_2^3}{2\,{\xi_1^3}}}\right) 
  \right]    
         \nonumber \\
&&\!\!\!\!\!\!\!\!\!\!
- 
 \left( 
     {\frac{63}{4}} + {\frac{31\,u}{2}} + 9\,{u^2} + 2\,{u^3} 
 \right) 
 \, {\rm Li}_3\left({\frac{\xi_1}{\xi_3}}\right) 
- 
 \left( 
   {\frac{127}{8}} + {\frac{55\,u}{4}} + {\frac{9\,{u^2}}{2}} +{u^3} 
 \right) \,
 \left[ 
   {\rm Li}_3\left({\frac{\xi_1^2\,\xi_3}{\xi_2\,\xi_1^3}}\right) 
 \right.
         \nonumber \\
&&\!\!\!\!\!\!\!\!\!\!
+\left. 
     {\rm Li}_3\left({\frac{\xi_1\,\xi_2^3}{\xi_2\,\xi_1^3}}\right) 
 \right]  
+  
 \left( 
  {\frac{7}{8}} + {\frac{15\,u}{4}} + {\frac{9\,{u^2}}{2}} + {u^3} 
 \right) \,
 \left[ 
     {\frac{1}{2}{\rm Li}_3
         \left(
         {\frac{(\xi_{-1}^{-2}+\xi_2) \,\xi_3}
           {{{({\xi_1^3})}^2}}}
         \right)} - 
     {\rm Li}_3\left({\frac{\xi_{-1}^{-2}+\xi_2}{{\xi_1^3}}}\right)
 \right.
         \nonumber \\
&&\!\!\!\!\!\!\!\!\!\! 
+ \left.
     {\rm Li}_3\left({\frac{\xi_1}{{\xi_1^3}}}\right) - 
     2\,{\rm Li}_3\left({\frac{{\xi_1^3}}{\xi_3}}\right) \
 \right]  
-                      
 \left( 
   {\frac{5}{2}} - {\frac{3\,u}{2}} - 6\,{u^2} - 2\,{u^3} 
 \right) \,
 {\rm Li}_3\left({\frac{\xi_{-1}}{\xi_2^3}}\right) 
 + 
 \left( {\frac{15}{4}} - 6\,u - 3\,{u^2} \right) \,
         \nonumber \\
&&\!\!\!\!\!\!\!\!\!\!
\times
 {\rm Li}_3\left({\frac{\xi_2}{\xi_2^3}}\right) 
 - 
 \left( 18 + 8\,u \right) \,
 {\rm Li}_3\left({\frac{\xi_2^3}{\xi_3}}\right)
+ 
 \left( 
  {\frac{119}{8}} + {\frac{47\,u}{4}} + {\frac{9\,{u^2}}{2}} +{u^3} 
 \right) \,
 \left[ 
   {\rm Li}_3\left({\frac{\xi_1^2}{{\xi_1^3}}}\right) + 
     {\rm Li}_3\left({\frac{\xi_2^3}{{\xi_1^3}}}\right) 
 \right] 
         \nonumber \\
&&\!\!\!\!\!\!\!\!\!\! 
 + 
 \left( {\frac{25}{16}} - {\frac{15\,u}{8}} - 
     {\frac{9\,{u^2}}{4}} - {\frac{{u^3}}{2}} 
 \right) 
 \, {\rm Li}_3\left(
       {\frac{\xi_1^{-2}\,\xi_1^2}{{{(\xi_2^3)}^2}}} 
              \right) \,.  
  \nonumber
\end{eqnarray}

\begin{eqnarray}
&&\!\!\!\!\!\!\!\!\!\! P2(u)
=     \\
&&\!\!\!\!\!\!\!\!\!\!
 4 
 \left( 1 + 4\,{u^2} \right) 
 \ln \left({\frac{\xi_{-1}}{\xi_1}}\right)
 \left[  
     {\rm Li}_2\left({\frac{\xi_1}{2}}\right)
      -{\rm Li}_2\left({\frac{-\xi_{-1}}{\xi_1}}\right)
 \right]  
 - 
 \left( 
  {\frac{33}{2}} + 59\,u + 62\,{u^2} + 20\,{u^3} 
 \right) \,
   \ln \left({\frac{\xi_1}{\xi_1^2}}\right)
       \nonumber   \\ 
&&\!\!\!\!\!\!\!\!\!\!
\times
   {\rm Li}_2\left(
        {\frac{2\,\xi_2}{{{(\xi_1^2)}^2}}}
            \right)
+\! 
  \left( 33 + 118\,u + 124\,{u^2} + 40\,{u^3} \right) 
   \ln \left({\frac{\xi_1}{\xi_1^2}}\right)
   {\rm Li}_2\left({\frac{2}{\xi_1^2}}\right) 
\!\!- \!\!
\left( {\frac{127}{8}} + {\frac{55\,u}{4}} + 
     {\frac{9\,{u^2}}{2}} + {u^3} 
\right) 
       \nonumber   \\ 
&&\!\!\!\!\!\!\!\!\!\!
\times
   \ln \left(
        {\frac{\xi_1\,\xi_2^3}{\xi_1^2\,\xi_3}}
       \right)
   {\rm Li}_2\left({\frac{\xi_1^2\,\xi_3}{\xi_2\,\xi_1^3}}\right) 
+ 
  \left( 
    {\frac{119}{8}} + {\frac{47\,u}{4}} + {\frac{9\,{u^2}}{2}}+{u^3} 
  \right) \,
   \ln \left({\frac{\xi_1^2}{\xi_2^3}}\right)\,
   {\rm Li}_2\left({\frac{\xi_2^3}{\xi_1^3}}\right) 
- 
  \Bigg[ 49 + 72\,u 
+    
    36\,{u^2}
       \nonumber   \\ 
&&\!\!\!\!\!\!\!\!\!\!
+ 
    \left( 
        {\frac{49}{4}} + 6\,u + 3\,{u^2} 
    \right) \,
      \ln (\xi_2)
+ 
    \left( 
      {\frac{37}{8}} + {\frac{39\,u}{4}} + 
        {\frac{3\,{u^2}}{2}} +  {u^3} 
    \right) \,
      \ln (\xi_{-1})  
+ 
    \Bigg( {\frac{145}{8}} + {\frac{425\,u}{4}} 
+  
        {\frac{239\,{u^2}}{2}} 
       \nonumber   \\ 
&&\!\!\!\!\!\!\!\!\!\!  
+  
    39\,{u^3} 
    \Bigg) \,
      \ln (\xi_1) 
-
    \left( {\frac{145}{8}} + {\frac{425\,u}{4}} + 
        {\frac{239\,{u^2}}{2}} + 39\,{u^3} 
    \right) \,
      \ln (\xi_1^2) 
- 
    \Bigg( {\frac{135}{8}} + {\frac{63\,u}{4}} + {\frac{9\,{u^2}}{2}} 
+ 
        {u^3} 
    \Bigg) \,
      \ln (\xi_2^3) 
  \Bigg] \,
       \nonumber   \\ 
&&\!\!\!\!\!\!\!\!\!\! 
 \times
   {\rm Li}_2\left({\frac{\xi_1^2}{\xi_2}}\right)
+ 
  \left( 
    {\frac{7}{16}} + {\frac{15\,u}{8}} + {\frac{9\,{u^2}}{4}} + 
     {\frac{{u^3}}{2}} 
  \right) \,
   \ln \left(
  {\frac{{{(\xi_1)}^2}}{\left( \xi_{-1}^{-2} + \xi_2 \right)\,\xi_3}}
        \right) \,
   {\rm Li}_2\left(
           {\frac{\left( \xi_{-1}^{-2} + \xi_2 \right)
                 \,\xi_3}{{{(\xi_1^3)}^2}}}
             \right) 
       \nonumber   \\ 
&&\!\!\!\!\!\!\!\!\!\!
+ 
    \xi_1\,\ln \left({\frac{\xi_{-1}^2}{\xi_2^3}}\right)\,
    {\rm Li}_2\left({\frac{\xi_2^3}{2\,\xi_2}}\right) 
- 
  \left( 
    {\frac{7}{8}} + {\frac{15\,u}{4}} + {\frac{9\,{u^2}}{2}} +{u^3} 
  \right) \,
    \ln \left({\frac{\xi_1}{\xi_{-1}^{-2} + \xi_2}}\right)\,
    {\rm Li}_2\left({\frac{\xi_{-1}^{-2} + \xi_2}{\xi_1^3}}\right)
       \nonumber   \\ 
&&\!\!\!\!\!\!\!\!\!\! 
+ 
  \frac{1}{8}
    \bigg\{
      \left( 25 - 30\,u - 36\,{u^2} - 8\,{u^3} \right) \, \ln3 
-    8\,
      \left( {\frac{15}{4}} - 6\,u - 3\,{u^2} \right) \,\ln (\xi_2) + 
       \xi_{-1}\,\xi_2^3\,
      \big[ 
         \xi_{-1}\,\ln (\xi_{-1}) 
       \nonumber   \\ 
&&\!\!\!\!\!\!\!\!\!\! 
+
  \xi_2^3\,\ln (\xi_1^2) 
      \big]  
    \bigg\} \,
     {\rm Li}_2\left({\frac{\xi_2}{\xi_2^3}}\right)
- 
  \bigg\{
    \frac{1}{2} \xi_{-1}^{-3} \xi_1 
    - (11 + 6\,u)\,\ln3 
- 
     \frac{1}{\xi_1^2} \,
    \left[ 
       \xi_{-1}^{-3} \, \xi_{-2}^{-3} 
     + \left({\xi_{-1}^{-2}}\right)^2 \,\ln3
    \right] 
       \nonumber   \\ 
&&\!\!\!\!\!\!\!\!\!\!
- 
    \left( 
       {\frac{105}{8}} + {\frac{17\,u}{4}} - 
        {\frac{9\,{u^2}}{2}} - {u^3} 
    \right) \,
     \ln (\xi_1) 
+ 
    \left( 
        {\frac{7}{8}} + {\frac{15\,u}{4}} + {\frac{9\,{u^2}}{2}} +
        {u^3} 
    \right) \,
     \ln (\xi_{-1}^{-2} + \xi_2)
       \nonumber   \\ 
&&\!\!\!\!\!\!\!\!\!\! 
+ 
    \left( {\frac{247}{8}} + {\frac{95\,u}{4}} + {\frac{9\,{u^2}}{2}} + 
        {u^3} 
    \right) \,
      \ln (\xi_1^2) 
- 
    \left( {\frac{135}{8}} + {\frac{63\,u}{4}} + {\frac{9\,{u^2}}{2}} + 
        {u^3} 
    \right) \,
      \ln (\xi_2^3) 
  \bigg\} \,
   {\rm Li}_2\left({\frac{\xi_1^3}{3\,\xi_1^2}}\right) 
       \nonumber   \\ 
&&\!\!\!\!\!\!\!\!\!\!
+ 
   \xi_1\,\ln \left({\frac{\xi_1}{3\,\xi_2^3}}\right)\,
   {\rm Li}_2\left({\frac{3\,\xi_2^3}{2\,\xi_1^3}}\right)
- 
  \bigg\{
     \left(  
        {\frac{95}{8}} + {\frac{39\,u}{4}} + {\frac{9\,{u^2}}{2}}+{u^3} 
     \right) \,
      \ln3 
+
     \frac{1}{\xi_1^2}
     \left[
       \xi_{-2} \, \xi_{-1} + \left( \xi_{-1}^{-2} \right)^2 \, \ln3
     \right]  
       \nonumber   \\ 
&&\!\!\!\!\!\!\!\!\!\!
-
     \frac{1}{2} \xi_{-2} \left( \xi_1^2 +\xi_{-2} \right)
- 
     \left( 
        {\frac{135}{8}} + {\frac{63\,u}{4}} + 
        {\frac{9\,{u^2}}{2}} + {u^3} 
     \right) \,
      \ln (\xi_1) 
+ 
     \left( 
       {\frac{127}{4}} + {\frac{55\,u}{2}} + 9\,{u^2} + 2\,{u^3} 
     \right) \,
      \ln (\xi_1^2) 
       \nonumber   \\ 
&&\!\!\!\!\!\!\!\!\!\!
- 
     \left( 
       {\frac{119}{8}} + {\frac{47\,u}{4}} + 
       {\frac{9\,{u^2}}{2}} +  {u^3} 
     \right) \,
       \ln (\xi_2^3) 
  \bigg\} \,
   {\rm Li}_2\left({\frac{\xi_2^3}{\xi_3}}\right) 
- 
  \left[ 
     2\,\xi_1\,\ln (\xi_1) - 2\,\xi_1\,\ln (\xi_2^3) 
  \right] \,
   {\rm Li}_2\left({\frac{\xi_2^3}{2\,\xi_1^2}}\right) \,.
   \nonumber
\end{eqnarray}

\begin{eqnarray}
&&\!\!\!\!\!\!\!\!\!\!  L3(u)
  =     \\
&&\!\!\!\!\!\!\!\!\!\!
 {\frac{1}{16}
 \left( 
    129 + 98\,u - 164\,{u^2} + 56\,{u^3} 
 \right) \,
   \ln (\xi_1)\,{{\ln (\xi_2)}^2}} 
 - 
  {\frac{1}{8}
 \left( 
   131 - 282\,u + 276\,{u^2} - 88\,{u^3} 
 \right) \,
   {{\ln (\xi_2)}^3}}
       \nonumber   \\ 
&&\!\!\!\!\!\!\!\!\!\!  
 + 
 {\frac{1}{16}
 \left( 
   295 - 798\,u + 804\,{u^2} - 264\,{u^3}  
 \right) \,
   {{\ln (\xi_1)}^2}\,\ln (\xi_1^2)} 
 + 
 \frac{1}{16}
 \left( 
     559 - 1742\,u + 1796\,{u^2} - 584\,{u^3} 
 \right) 
       \nonumber   \\ 
&&\!\!\!\!\!\!\!\!\!\!
\times
   {{\ln (\xi_2)}^2}\,\ln (\xi_1^2) 
 - 
 \left( 
 {\frac{49}{2}} - 65\,u + 62\,{u^2} - 20\,{u^3} 
 \right) \,
   \ln (\xi_1)\,{{\ln (\xi_1^2)}^2} 
 - 
 \left( 
    33 - 118\,u + 124\,{u^2} - 40\,{u^3} 
 \right) 
       \nonumber   \\ 
&&\!\!\!\!\!\!\!\!\!\!
\times
   \ln (\xi_2)\,{{\ln (\xi_1^2)}^2} 
 - 
 {\frac{1}{8}
   {{(\xi_1)}^2}\,(\xi_{-1}^{-2}+\xi_2)\,
   {{\ln (\xi_1)}^2}\,\ln (\xi_1^3)}
 + 
 \frac{1}{48}
 \left( 
   1087 - 3054\,u + 3012\,{u^2} - 968\,{u^3} 
 \right) 
       \nonumber   \\ 
&&\!\!\!\!\!\!\!\!\!\!
\times
   {{\ln (\xi_1^2)}^3}  
 - 
 {\frac{1}{16}
 \left( 
      89 - 142\,u + 60\,{u^2} - 8\,{u^3} 
 \right) \,
      {{\ln (\xi_2)}^2}\,\ln (\xi_2^3)} 
 +
  \frac{1}{8}
  \left( 
      119 - 94\,u + 36\,{u^2} - 8\,{u^3} 
  \right) 
       \nonumber   \\ 
&&\!\!\!\!\!\!\!\!\!\!
\times
  \left[
    \ln (\xi_1^2)\,{{\ln (\xi_1^3)}^2} 
 + 
    \ln (\xi_1)\,\ln (\xi_2)\, \ln (\xi_2^3) 
 + 
    \ln (\xi_1)\,\ln (\xi_1^3)\, \ln (\xi_2^3)
 - 
    2\, \ln (\xi_1)\,\ln (\xi_1^2)\, \ln (\xi_2^3)
  \right]
       \nonumber   \\ 
&&\!\!\!\!\!\!\!\!\!\!
 + 
 \left( 
   {\frac{183}{4}} - {\frac{71\,u}{2}} + 9\,{u^2} - 2\,{u^3} 
 \right) \,
   \ln (\xi_1)\,\ln (\xi_1^2)\,\ln (\xi_1^3) 
 - 
 \frac{1}{8}
 \left( 
   105 - 34\,u - 36\,{u^2} + 8\,{u^3} 
 \right) 
      \ln (\xi_1)\,{{\ln (\xi_1^3)}^2} 
       \nonumber   \\ 
&&\!\!\!\!\!\!\!\!\!\!
 - 
 {\frac{1}{8}
   {{(\xi_2^3)}^2}\, \xi_{-1}\,
   \ln (\xi_2)\,\ln (\xi_1^2)\, \ln (\xi_2^3)}
 - 
 {\frac{1}{12} {{(\xi_1)}^2}\,(\xi_{-1}^{-2}+\xi_2)\,
      {{\ln (\xi_1^3)}^3}}  
 + 
 \frac{1}{8}
 \left(
    127 - 110\,u + 36\,{u^2} - 8\,{u^3} 
 \right) 
       \nonumber   \\ 
&&\!\!\!\!\!\!\!\!\!\!
\times
 \left[
   {{\ln (\xi_2)}^2}\, \ln (\xi_1^3)
 +
   \ln (\xi_2)\,{{\ln (\xi_1^3)}^2} 
 - 
   2\, \ln (\xi_1)\,\ln (\xi_2)\, \ln (\xi_1^3)
 -
   {{\ln (\xi_1^3)}^2}\,\ln (\xi_2^3) 
 + 
   \ln (\xi_2\,\xi_1^3)\,{{\ln (\xi_2^3)}^2}
 \right]
       \nonumber   \\ 
&&\!\!\!\!\!\!\!\!\!\!
 +
 {\frac{1}{8}
 \left( 
      135 - 126\,u + 36\,{u^2} - 8\,{u^3} 
 \right) \,
   \ln (\xi_1^2)\,\ln (\xi_1^3)\, \ln (\xi_2^3)} 
 - 
 \frac{1}{16}
 \left( 
     433 - 530\,u + 252\,{u^2} - 56\,{u^3} 
 \right) 
       \nonumber   \\ 
&&\!\!\!\!\!\!\!\!\!\!
\times
   \ln (\xi_1^2)\,{{\ln (\xi_2^3)}^2} 
 -
 {\frac{1}{24}
 \left( 
      55 + 78\,u - 60\,{u^2} + 8\,{u^3} 
 \right) \,
   {{\ln (\xi_2^3)}^3}}
 + 
 \frac{1}{8}
 \left( 
      293 - 266\,u + 108\,{u^2} - 24\,{u^3} 
 \right) 
       \nonumber   \\ 
&&\!\!\!\!\!\!\!\!\!\!
\times
   {{\ln (\xi_1^2)}^2}\,\ln (\xi_2^3)  
 - 
 {\frac{1}{16}
 \left( 
     69 - 18\,u + 140\,{u^2} + 264\,{u^3} 
 \right) \,
   {{\ln (\xi_1)}^2}\,\ln (\xi_{-1})} 
 + 
 \left( 
  {\frac{37}{8}} - {\frac{39\,u}{4}} + {\frac{3\,{u^2}}{2}} -{u^3} 
 \right) 
       \nonumber   \\ 
&&\!\!\!\!\!\!\!\!\!\!
\times
   \ln (\xi_1)\,\ln (\xi_2)\, \ln (\xi_{-1}) 
 - 
 {\frac{1}{3}
 \left( 
    10 + 48\,u + 24\,{u^2} + 32\,{u^3} 
 \right) \,
 {{\ln (\xi_{-1})}^3}} 
 + 
 {\frac{1}{8}
   {{(\xi_1)}^2}\,\xi_{-2}^{-3}\,
   {{\ln (\xi_1)}^2}\,\ln (\xi_{-2})} 
       \nonumber   \\ 
&&\!\!\!\!\!\!\!\!\!\!
 - 
 {\frac{1}{16}
 \left( 
    67 + 30\,u - 12\,{u^2} + 8\,{u^3} 
 \right) \,
   \ln (\xi_1)\,{{\ln (\xi_{-2})}^2}} 
 - 
 {\frac{1}{8}
    {{(\xi_1)}^2}\,(\xi_{-1}^{-2}+\xi_2)\,
    \ln (\xi_1)\,\ln (\xi_1^3)\, \ln (\xi_{-1}^{-2}+\xi_2)} 
       \nonumber   \\ 
&&\!\!\!\!\!\!\!\!\!\!
 - 
 \frac{1}{8}
 \left( 
     373 - 314\,u + 108\,{u^2} - 24\,{u^3} 
 \right) 
   {{\ln (\xi_1^2)}^2}\,\ln (\xi_1^3) 
 + 
 {\frac{1}{8}
    {{(\xi_1)}^2}\,(\xi_{-1}^{-2}+\xi_2)\,
    \ln (\xi_1^2)\,\ln (\xi_1^3)\, \ln (\xi_{-1}^{-2}+\xi_2)} 
       \nonumber   \\ 
&&\!\!\!\!\!\!\!\!\!\!
 - 
 {\frac{1}{8}
 \left( 
    31 + 146\,u - 188\,{u^2} + 56\,{u^3} 
 \right) 
   \ln (\xi_1)\,\ln (\xi_2)\, \ln (\xi_1^2)} 
 + 
 \frac{1}{16}
  \left( 
    35 - 66\,u - 12\,{u^2} + 8\,{u^3} 
  \right) 
  \ln (\xi_1) 
       \nonumber   \\ 
&&\!\!\!\!\!\!\!\!\!\
\times
 {{\ln (\xi_{-2}^{-3})}^2}
 \! +\! 
 {\frac{1}{8}
   \xi_1\,{{(\xi_{-2}^{-3})}^2}
   \ln (\xi_1)  \ln (\xi_{-2}) \ln (\xi_{-1}^{-2})} 
 \!+\! 
 {\frac{1}{16}
 {{(\xi_1)}^2} (\xi_{-1}^{-2}+\xi_2)
 \left[ 
    {{\ln (\xi_1)}^2} \!-\! {{\ln (\xi_1^2)}^2} 
 \right] 
   \ln (\xi_{-1}^{-2}+\xi_2)}
       \nonumber   \\ 
&&\!\!\!\!\!\!\!\!\!\!
 - \!
 {\frac{1}{8}
   \xi_1\,{{(\xi_{-2}^{-3})}^2}\,\ln (\xi_1)
   \ln (\xi_{-1}^{-2}) \ln (\xi_{-2}^{-3})} 
 \!-\! 
 {\frac{1}{8}
   {{(\xi_1)}^2} \xi_{-2}^{-3} \ln (\xi_1)
   \ln (\xi_{-2}) \ln (\xi_{-2}^{-3})} 
 \!-\! 
 {\frac{1}{16}
   {{(\xi_1)}^2}\,\xi_{-2}^{-3}
   {{\ln (\xi_1)}^2} \ln (\xi_{-2}^{-3})} \,.
  \nonumber
\end{eqnarray}

\begin{eqnarray}
&&\!\!\!\!\!\!\!\!\!\!  L2(u)
=     \\
&&\!\!\!\!\!\!\!\!\!\!
 \frac{-1}{\xi_1^2\,\xi_{-1}} \,
 \left[
    49 + 12\,\ln2  
    - u\,
    \left( 
       38 - 16\,\ln2 
    \right) 
    - 4\,{u^2}\,
    \left( 
       45 - 8\,\ln 2 
    \right)  
    + 8\,{u^3}\,
    \left( 
      15 + 8\,\ln2 
    \right) 
    - 64\,{u^4}\,\ln 2
 \right]
       \nonumber   \\ 
&&\!\!\!\!\!\!\!\!\!\!
\times
  {{\ln (\xi_1)}^2} 
 + 
  \big[ 
    49 + 2\,\ln2 - 2\,\ln3 
    + 4\,u\,\left( -18 - \ln2 + \ln3 \right) 
    + 36\,{u^2}  
  \big] \,
  \ln (\xi_1)\,\ln (\xi_2)
       \nonumber   \\ 
&&\!\!\!\!\!\!\!\!\!\! 
 -
 \frac{1}{2 \xi_1^2 \, \xi_1} \,
 \big[  
   127  + 6\,\ln2 - 6\,\ln3  
   - 4\,u\,\left( 123 + 7\,\ln2 - 7\,\ln3 \right) 
   + 8\,{u^2}\,( 92 + 5\,\ln2 - 5\,\ln3 ) 
       \nonumber   \\ 
&&\!\!\!\!\!\!\!\!\!\! 
   - 16\,{u^3}\, \left( 33  \ln2 - \ln3 \right)   
   + 144\,{u^4} 
 \big]
 {{\ln (\xi_2)}^2} 
 + 
  {\frac{2}{\xi_1^2} \,
  \left[ 
    -19 + 35\,u - 20\,{u^2} + 4\,{u^3} 
  \right]  
  \ln (\xi_2)\,\ln (\xi_1^2)} 
       \nonumber   \\ 
&&\!\!\!\!\!\!\!\!\!\!
 - 
 \frac{2}{\xi_1^2} \,
 \left[ 
    14 + 6\,\ln2 + 18\,\ln3 
    - u\,\left( 23 + 16\,\ln2 + 18\,\ln3 \right)  
    + 4\,{u^2}\,\left( 4 + 2\,\ln2 + \ln3 \right)  
    - 4\,{u^3} 
 \right]
       \nonumber   \\ 
&&\!\!\!\!\!\!\!\!\!\!
\times
    \ln (\xi_1)\,\ln (\xi_1^2) 
 - 
 \xi_1\,\ln2\,{{\ln (\xi_1^3)}^2} 
 + 
 \frac{1}{16}
 \big[ 
    176 + 32\,\ln2 + 263\,\ln3 
    - 2\,u\,\left( 96 + 32\,\ln2 + 79\,\ln3 \right)
       \nonumber   \\ 
&&\!\!\!\!\!\!\!\!\!\!
    + 4\,{u^2}\,\left( 16 + 9\,\ln3 \right) 
    - 8\,{u^3}\,\ln3    
 \big] \,
 {{\ln (\xi_1^2)}^2} 
+ 
 \frac{1}{8}
 \big[ 
    16\,\ln2 + 231\,\ln3 
    - 2\, u\, (16\, \ln2 + 79\, \ln3) 
       \nonumber   \\ 
&&\!\!\!\!\!\!\!\!\!\!
    + 36\,{u^2}\,\ln3 
    - 8\,{u^3}\,\ln3 \
 \big] \,
 \ln (\xi_1)\,\ln (\xi_1^3) 
 - 
 \frac{1}{8}
 \big[ 
      119 - 94\,u + 36\,{u^2} - 8\,{u^3} 
 \big] \,\ln3\,
 \ln (\xi_1)\,\ln (\xi_2^3) 
       \nonumber   \\ 
&&\!\!\!\!\!\!\!\!\!\!
 -
 \frac{1}{8}
 \big[ 
    373
    - 314\,u 
    + 108\,{u^2} 
    - 24\,{u^3}
 \big] \,
 \ln3 \, \ln (\xi_1^2)\,\ln (\xi_1^3) 
 + \frac{1}{8}
 \big[ 
    213 
    - 218\,u 
    + 108\,{u^2} 
    - 24\,{u^3} 
 \big] \,\ln3\,
       \nonumber   \\ 
&&\!\!\!\!\!\!\!\!\!\!
\times
      \ln (\xi_1^2) \ln (\xi_2^3) 
 - 
 {\frac{1}{8}
    {{(\xi_2^3)}^2}\,\xi_{-1}\,\ln3
    \ln (\xi_2) \ln (\xi_2^3)} 
 + 
 \frac{1}{8}
 ( 
   135
   - 126\,u + 36\,{u^2} - 8\,{u^3} 
 ) \ln3 
      \ln (\xi_1^3)\ln (\xi_2^3)
       \nonumber   \\ 
&&\!\!\!\!\!\!\!\!\!\!
 - 
 \frac{1}{16}
 ( 
   179 - 310\,u + 180\,{u^2} 
   - 40\,{u^3} 
 ) \,\ln3\,
      {{\ln (\xi_2^3)}^2}
 - 
 \frac{1}{2}
 \big[ 
   13 - 8\,\ln2 
   - 39\,u\,\ln2 
   + {u^2}\,\left( 36 - 32\,\ln2 \right) 
       \nonumber   \\ 
&&\!\!\!\!\!\!\!\!\!\!
   - 4\,{u^3}\,\ln2 
 \big] \,
 \ln (\xi_1)\,\ln (\xi_{-1}) 
 + 
 \frac{1}{8\,\xi_{-1}}\,
 \big[
        48 + 25\,\ln3   
        +  4\,u \,\left( 12 + 5\,\ln3 \right)  
        +  32\,{u^2}\,\left( 2 - 3\,\ln3 \right)
       \nonumber   \\ 
&&\!\!\!\!\!\!\!\!\!\!  
        +  16\,{u^3}\,\left( 4 - 5\,\ln3 \right)
        - 16\,{u^4}\,\ln3 
 \big] 
  \ln (\xi_1)\,\ln (\xi_{-2}) 
 - 
 {\frac{1}{8}
    {{(\xi_1)}^2}\,(\xi_{-1}^{-2}+\xi_2)\,\ln3\,
      \ln \left({\frac{\xi_1^2}{\xi_1^3}} \right)\,
      \ln (\xi_{-1}^{-2}+\xi_2)} 
       \nonumber   \\ 
&&\!\!\!\!\!\!\!\!\!\!
- 
  {\frac{1}{8}
     \xi_1\,{{(\xi_{-2}^{-3})}^2}\,\ln3\,
      \ln (\xi_1)\,\ln (\xi_{-2}^{-3})} \, .
   \nonumber
\end{eqnarray}

\begin{eqnarray}
&&\!\!\!\!\!\!\!\!\!\!  L1(u)
=     \\
&&\!\!\!\!\!\!\!\!\!\!
 \left[
  6\,\xi_2   + 
  \left(295 - 798\,u + 804\,u^2 - 264\,u^3 \right) \,
  \frac{{\pi }^2}{48}
 - 
  \xi_1  \,  {\ln \left(\frac{3}{2}\right)}^2 
 - 
  \left( 19 - 16\,u + 4\,u^2 \right) \,\ln3 \,
  \frac{\xi_2}{2\, \xi_1^2}
 \right]
 \ln (\xi_2) 
       \nonumber   \\ 
&&\!\!\!\!\!\!\!\!\!\!
 -
 \bigg[ 
     \left( 79 - 402\,u + 356\,u^2 - 376\,u^3 \right) \,
  \frac{{\pi }^2}{48} 
 + 
     \left( 7 - 8\,u + 4\,u^2 \right) \,\ln3 \,
     \frac{\xi_1^3\,}{2\, \xi_1^2} 
 + 
   2\,\left( 3 - u \right) \,
   {\ln (3)}^2 
       \nonumber   \\ 
&&\!\!\!\!\!\!\!\!\!\!
 + 
   \xi_1\, \left( 15 + 2\,\ln2\,\ln3 \right)
 \bigg]
   \ln (\xi_1)  
 -
 \bigg[
  3\,\xi_1^2 
 + \left( 861 - 1354\,u + 
       1100\,u^2 - 344\,u^3 
   \right) \, \frac{{\pi }^2}{96} 
 - 
  2\,\xi_1\,{\ln (2)}^2 
       \nonumber   \\ 
&&\!\!\!\!\!\!\!\!\!\!
 + 
 \frac{1}{48}
 \bigg[
   \left( 197 + 22\,u + 12\,u^2 - 24\,u^3 \right) \,{\pi }^2 + 
  6\,\left( 47 - 62\,u + 36\,u^2 - 8\,u^3 \right) \,
   {\ln (3)}^2
 \bigg] \,
  \ln (\xi_2^3)
       \nonumber   \\ 
&&\!\!\!\!\!\!\!\!\!\!
 -
 \bigg[ 
 \xi_1\,{\ln (2)}^2 
 -
    {(\xi_1)}^2\,(\xi_{-1}^{-2}+\xi_2)\frac{{\pi }^2}{24}
 +
  \left( \frac{119}{8} - \frac{47\,u}{4} +
     \frac{9\,u^2}{2} - u^3 \right) \,{\ln (3)}^2
 \bigg] \,
 \ln (\xi_1^3)
       \nonumber   \\ 
&&\!\!\!\!\!\!\!\!\!\!
 - 
  2\,\left( 11 - 12\,u + 4\,u^2 \right) \,\ln3 
 - 
  \frac{1}{16}
  \left( 263 - 158\,u + 36\,u^2 - 8\,u^3 \right) 
   \,{\ln (3)}^2
 \bigg] \,
 \ln (\xi_1^2)
       \nonumber   \\ 
&&\!\!\!\!\!\!\!\!\!\! 
 + 
 {\frac{1}{96}
 {{(\xi_1)}^2}\,(\xi_{-1}^{-2}+\xi_2)\,
 \left[ {{\pi }^2} - 6\,{{\ln (3)}^2} 
 \right] \,
 \ln (\xi_{-1}^{-2}+\xi_2)}  \,.
   \nonumber
\end{eqnarray}

\begin{eqnarray}
&&\!\!\!\!\!\!\!\!\!\!  R(u)
=     \\
&&\!\!\!\!\!\!\!\!\!\!
 -{\frac{{{\pi }^2}}{2}} 
 + 6\,{{\pi }^2}\,\ln2 
 - 48\,{{\ln (2)}^3} 
 - 
 \left[
   9 - {\frac{265\,{{\pi }^2}}{96}} - {{\ln (2)}^2}
 \right] \, \ln3 
 + 
 \left(
 11 + 37\,\ln2 
 \right)\,{{\ln (3)}^2} 
 -
 \frac{953}{48} \, {\ln(3)}^3
       \nonumber   \\ 
&&\!\!\!\!\!\!\!\!\!\! 
 - 
  38\,\ln3\,{\rm Li}_2\left({\frac{2}{3}}\right) - 
  18\,\ln3\,{\rm Li}_2\left({\frac{3}{4}}\right) + 
  18\,{\rm Li}_3\left({\frac{1}{4}}\right) 
 + 
  38\,{\rm Li}_3\left({\frac{1}{3}}\right) + 
  18\,{\rm Li}_3\left({\frac{3}{4}}\right) 
 - {\frac{1845}{16}}\,\zeta(3)
       \nonumber   \\ 
&&\!\!\!\!\!\!\!\!\!\! 
 + 
  {u^2}\,
  \left[ {\frac{4\,{{\pi }^2}}{3}} - 
     {\frac{3\,{{\pi }^2}}{8}}\,\ln3 + 4\,{{\ln (3)}^2} + 
     {\frac{3}{4}}\,{{\ln (3)}^3} - {\frac{651}{4}}\,\zeta(3) \
  \right]  \, .
   \nonumber
\end{eqnarray}

\end{appendix}

\raggedcolumns                                        %
\begin{multicols}{2}                                  %
\narrowtext                                           %


\begin{thebibliography}{10}

\bibitem{Nato92}
{\em Single Charge Tunneling}, Vol.~294 of {\em NATO ASI Series B}, edited by
  H. Grabert and M.~H. Devoret (Plenum, New York, 1992).

\bibitem{Averin91}
D.~V. Averin and K.~K. Likharev,  in {\em Mesoscopic Phenomena in Solids},
  Vol.~30 of {\em Modern Problems in Condensed Matter Science}, edited by B.~L.
  Altshuler, P.~A. Lee, and R.~A. Webb (North-Holland, Amsterdam, 1991), p.\
  173.

\bibitem{IngoldNato92in}
G.~L. Ingold and Y.~V. Nazarov, in Ref.~\protect\cite{Nato92}, p. 21.

\bibitem{AverinNato92in}
D.~V. Averin and Y.~V. Nazarov, in Ref.~\protect\cite{Nato92}, p. 217.

\bibitem{MatveevBOXJETP91}
K.~A. Matveev, Sov.\ Phys.\ JETP {\bf 72},  892  (1991).

\bibitem{Schoeller2StatePRB94}
H. Schoeller and G. Sch\"on, Phys.\ Rev.\ B {\bf 50},  18436  (1994).

\bibitem{ZaikinNCAPRB94}
D.~S. Golubev and A.~D. Zaikin, Phys.\ Rev.\ B {\bf 50},  8736  (1994).

\bibitem{FalciScalePRL95}
G. Falci, G. Sch\"on, and G.~T. Zimanyi, Phys.\ Rev.\ Lett.\ {\bf 74},  3257
  (1995).

\bibitem{KoenigRGPRL98}
J. K\"onig and H. Schoeller, Phys.\ Rev.\ Lett.\ {\bf 81},  3511  (1998).

\bibitem{SchoenREP90}
G. Sch\"on and A.~D. Zaikin, Phys.\ Rep.\ {\bf 198},  237  (1990).

\bibitem{GeorgSemiclEPJB00}
G. G\"oppert and H. Grabert, Eur.\ Phys.\ J.\ B {\bf 16}, 687 (2000).

\bibitem{BenJacPRL83}
E. Ben-Jacob, E. Mottola, and G. Sch\"on, Phys.\ Rev.\ Lett.\ {\bf 51},  2064
  (1983).

\bibitem{ZaikinECPRL91}
S.~V. Panyukov and A.~D. Zaikin, Phys.\ Rev.\ Lett.\ {\bf 67},  3168  (1991).

\bibitem{ZaikinSJPRB92}
D.~S. Golubev and A.~D. Zaikin, Phys.\ Rev.\ B {\bf 46},  10903  (1992).

\bibitem{ZaikinSETJETP96}
D.~S. Golubev and A.~D. Zaikin, JETP Lett.\ {\bf 63},  1007  (1996).

\bibitem{WangBOXPRB96}
X. Wang and H. Grabert, Phys.\ Rev.\ B {\bf 53},  12621  (1996).

\bibitem{GeorgSJPRB97}
G. G\"oppert, X. Wang, and H. Grabert, Phys.\ Rev.\ B {\bf 55},  R10213
  (1997).

\bibitem{GeorgSETPRB98}
G. G\"oppert and H. Grabert, Phys.\ Rev.\ B {\bf 58},  R10155  (1998).

\bibitem{GeorgSEMICRAS99}
G. G\"oppert and H. Grabert, C.\ R.\ Acad.\ Sci.\ {\bf 327},  885  (1999).

\bibitem{WangMCEPL97}
X. Wang, R. Egger, and H. Grabert, Europhys.\ Lett.\ {\bf 38},  545  (1997).

\bibitem{ZwergerBOXPRL97}
W. Hofstetter and W. Zwerger, Phys.\ Rev.\ Lett.\ {\bf 78},  3737  (1997).

\bibitem{HerreroMCPRB99}
C.~P. Herrero, G. Sch\"on, and A.~D. Zaikin, Phys.\ Rev.\ B {\bf 59},  5728
  (1999).

\bibitem{GeorgSVDPhysicaB00}
G. G\"oppert, B. H\"upper, and H. Grabert, Physica B {\bf 284-288},  1792
  (2000).

\bibitem{GrabertBOXPRB94}
H. Grabert, Phys.\ Rev.\ B {\bf 50},  17364  (1994).

\bibitem{GeorgBOXPRL98}
G. G\"oppert, H. Grabert, N.~V. Prokof'ev, and B.~V. Svistunov, Phys.\ Rev.\
  Lett.\ {\bf 81},  2324  (1998).

\bibitem{KoenigRENPRB98}
J. K\"onig, H. Schoeller, and G. Sch\"on, Phys.\ Rev.\ B {\bf 58},  7882
  (1998).

\bibitem{LafargeZPB91}
P. Lafarge {\it et~al.}, Z.\ Phys.\ B {\bf 85},  327  (1991).

\bibitem{GeorgChannelEPL99}
G. G\"oppert, H. Grabert, and C. Beck, Europhys.\ Lett.\ {\bf 45},  249
  (1999).

\bibitem{Lewin81}
L. Lewin, {\em Polylogarithms and Associated Functions} (North Holland, New
  York, 1981).

\bibitem{EsteveNato92in}
D. Esteve, in Ref.~\protect\cite{Nato92}, p. 109.

\bibitem{GrabertBOXPhysica94}
H. Grabert, Physica B {\bf 194-196},  1011  (1994).

\bibitem{GrabertPOT82}
H. Grabert, {\em Projection Operator Techniques in Nonequilibrium Statistical
  Mechanics} (Springer, New York, 1982).

\bibitem{SchoenPRB85}
G. Sch\"on, Phys.\ Rev.\ B {\bf 32},  4469  (1985).

\bibitem{JoyezSETPRL97}
P. Joyez {\it et~al.}, Phys.\ Rev.\ Lett.\ {\bf 79},  1349  (1997).

\bibitem{KoenigCOTPRL97}
J. K\"onig, H. Schoeller, and G. Sch\"on, Phys.\ Rev.\ Lett.\ {\bf 78},  4482
  (1997).

\end{thebibliography}

\end{multicols}                                       %

\end{document}